\documentclass[aps,prb,twocolumn,preprintnumbers,amsmath,amssymb,superscriptaddress]{revtex4-2}

\usepackage{amsmath, amssymb}
\usepackage{mathtools}
\usepackage{physics}
\usepackage{graphicx}
\usepackage{hyperref}
\usepackage{slashed}
\usepackage{orcidlink}
\usepackage{bbm}
\usepackage{mathtools}
\usepackage{newtxtext}
\usepackage[varvw]{newtxmath}
\usepackage{placeins}
\usepackage{bbm}
\usepackage[normalem]{ulem}

\hypersetup{
    colorlinks,
    linkcolor={blue},
    citecolor={blue},
    urlcolor={blue}
}

\newcommand{\tdbWSe}{tdbWSe$_2$}

\begin{document}

\title{Relativistic Mott transition and high-order van Hove singularity in twisted double bilayer WSe${}_2$:\\
mean-field and functional renormalization group study}

\author{Bilal Hawashin}
\email{hawashin@tp3.rub.de}
\affiliation{Theoretische Physik III, Ruhr-Universit\"at Bochum, D-44801 Bochum, Germany}

\author{Julian Kleeschulte}
\email{julian.kleeschulte@rub.de}
\affiliation{Theoretische Physik III, Ruhr-Universit\"at Bochum, D-44801 Bochum, Germany}

\author{David Kurz}
\email{david.kurz@rub.de}
\affiliation{Theoretische Physik III, Ruhr-Universit\"at Bochum, D-44801 Bochum, Germany}

\author{Aiman Al-Eryani}
\email{aiman.al-eryani@rub.de}
\affiliation{Theoretische Physik III, Ruhr-Universit\"at Bochum, D-44801 Bochum, Germany}

\author{Michael M. Scherer}
\email{scherer@tp3.rub.de}
\affiliation{Theoretische Physik III, Ruhr-Universit\"at Bochum, D-44801 Bochum, Germany}

\begin{abstract}
Experiments on twisted double bilayer tungsten diselenide have demonstrated that moir'e semiconductors can realize a relativistic Mott transition, i.e., a quantum phase transition from a Dirac semimetal to a correlated insulating state, by twist-angle tuning.
In addition, signatures of van Hove singularities were observed in the material's moir'e valence bands, suggesting further potential for the emergence of strongly-correlated states.
Based on a continuum model, we provide a detailed analysis of the twist-angle dependence of the system's band structure, focusing on the evolution of the Dirac excitations and the Fermi-surface structure with its Lifshitz transitions across the van Hove fillings.
We exhibit that the twist angle can be used to band engineer a high-order van Hove singularity, which can be accessed by gate tuning.
We then study the magnetic phase diagram of an effective Hubbard model for twisted double bilayer tungsten diselenide on the effective honeycomb superlattice with tight-binding parameters fitted to the two topmost bands of the continuum model.
To that end, we employ a self-consistent Hartree-Fock mean-field approach in real space.
Fixing the angle-dependent Hubbard interaction based on the experimental findings, we explore a broad parameter range of twist angle, filling, and temperature.
We find a rich variety of magnetic states that we expect to be accessible in future experiments, including, e.g., a non-coplanar spin-density wave with non-zero spin chirality and a half-metallic uniaxial spin-density wave.
Finally, we employ a functional renormalization group approach to also study the competition between density-wave and superconducting instabilities.
For twist angles of $\theta=2.0^\circ, 2.5^\circ$, as well as $\theta\approx 3.5^\circ$ -- where the high-order van Hove-singularity is found -- we find clear indications for unconventional superconductivity.
\end{abstract}
\maketitle

\section{Introduction} 

An exciting correlation effect that was discussed to occur in single-layer graphene, due to the presence of gapless Dirac excitations and sizable Coulomb interactions, is the relativistic Mott transition~\cite{PhysRevLett.97.146401,PhysRevB.79.085116,PhysRevB.80.075432,PhysRevB.80.205319,Semenoff_2012,Boyack:2020xpe,Herbut:2023xgz}: 
strong electron-electron interactions may break the chiral symmetry spontaneously and generate a finite mass gap for the Dirac excitations. 
While this phenomenon would be a beautiful analogy to dynamical mass generation in elementary particle physics, it was never observed in pristine graphene, presumably due to the electron-electron interactions not being strong enough.

Two-dimensional van der Waals materials, however, are a highly tunable platform for engineering electronic band structures and interactions~\cite{doi:10.1126/science.aac9439,Andrei_2020,mak2022semiconductor} and have been put forward as quantum simulators of strongly correlated and topological physics~\cite{Kennes_2021}.
Indeed, recently, the relativistic Mott transition was observed in ``artificial graphene'' made from twisted bilayers of WSe${}_2$ bilayers of AB structure~\cite{ma2024relativistic} upon small twist-angle tuning starting from 180${}^\circ$ (ABBA stacking).
More precisely, the transport data in Ref.~\cite{ma2024relativistic} shows that the system's $\Gamma$~valley moir\'e valence bands emulate graphene's characteristic massless Dirac fermions at half filling as well as its van Hove singularities (VHS).
They further report that for small twists below $\sim 2.7^\circ$ the system turns from a semimetal into an insulator as distinctive for a relativistic Mott transition.
Such a transition is facilitated by the systematic suppression of the Fermi velocity at small angles, boosting interaction effects. 
This is corroborated by band-structure calculations for twisted double bilayer WSe${}_2$ (\tdbWSe) in Refs.~\cite{Pan:2023,Biedermann:2025dma}, where it was argued in a mean-field approach that a N\'eel antiferromagnet is a good candidate for the insulating state at Dirac filling.

Interestingly, at the van Hove fillings -- where due to the high density of states (DOS) a tendency towards the formation of correlated states could be expected -- the system remains metallic, at least slightly above and below the critical angle for the relativistic Mott transition~\cite{ma2024relativistic}.
Related experiments on the formation of insulating behavior in tunable Dirac systems have been reported on in twisted bilayer molybdenum diselenide~\cite{yang2025correlated} and theory proposals for the observation of the relativistic Mott transition in twisted bilayer graphene can be found in Refs.~\cite{hj61-dw78,Huang_2025}.

In this work, we provide an extended analysis of the moir\'e bands and Fermi-surface structure of \tdbWSe, and  perform unrestricted Hartree-Fock mean-field and functional renormalization group calculations to determine the dominant order for the full range of experimentally explored twist angles and fillings.
To that end, we start from the continuum-model approach put forward in Ref.~\cite{Pan:2023} and finely resolve the band structure for $\theta\in [2.0^\circ,4.0^\circ]$.
One key finding is that each saddle point in the energy dispersion of the second-to-highest moir\'e band splits into two, when increasing the twist angle beyond $\theta_c\sim 3.58^\circ$.
Right at~$\theta_c$, the corresponding VHS is then of high order, featuring a power-law divergent DOS instead of the logarithmic one at a conventional saddle point point.

Next, we fit the two topmost moir\'e bands from the continuum model to a honeycomb-lattice tight-binding model with hopping amplitudes up to the tenth-nearest neighbor.
We then add an on-site interaction and perform unrestricted Hartree-Fock mean-field calculations in real space for the resulting Hubbard model.
To that end, we choose the Hubbard interaction such that it induces a relativistic Mott transition at the observed twist of $\sim 2.7^\circ$, and use a scaling property to obtain interaction parameters for the full range of  angles~\cite{Wu:2018}.  
We present phase diagrams of the emerging magnetic states for twist angles $\theta\in [2.0^\circ,4.0^\circ]$ and in a broad range of fillings stretching from below van Hove filling in the valence band to above van Hove filling in the conduction band.

Finally, we employ a functional renormalization group (fRG) approach to explore the competition of density-wave instabilities, including the tendencies towards magnetic order, and superconducting instabilities.

\begin{figure*}
    \centering
    \includegraphics[scale=0.39]{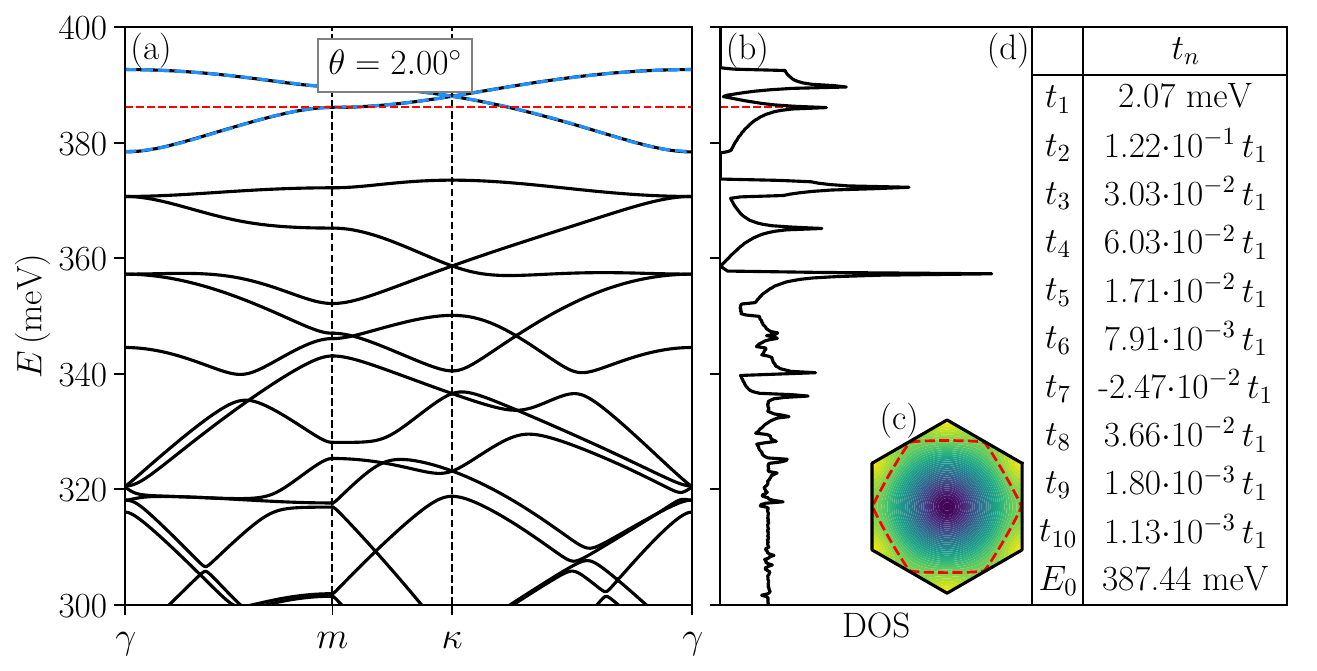}
    \includegraphics[scale=0.39]{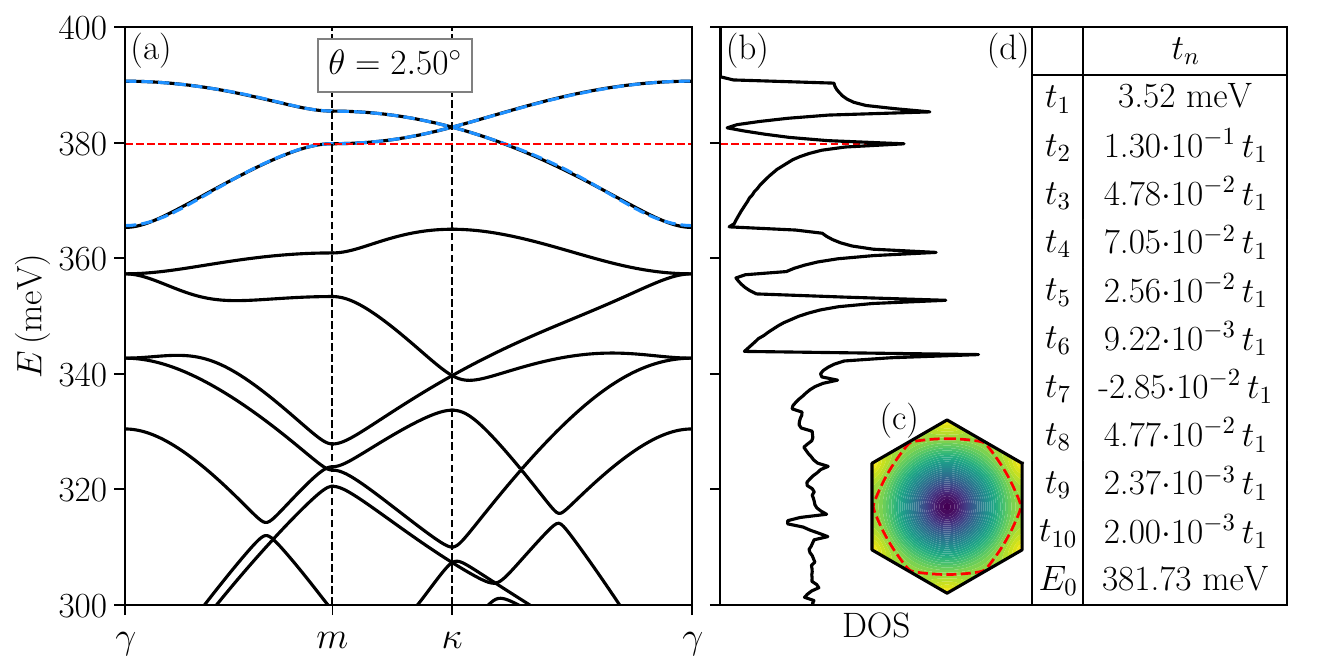}
    \includegraphics[scale=0.39]{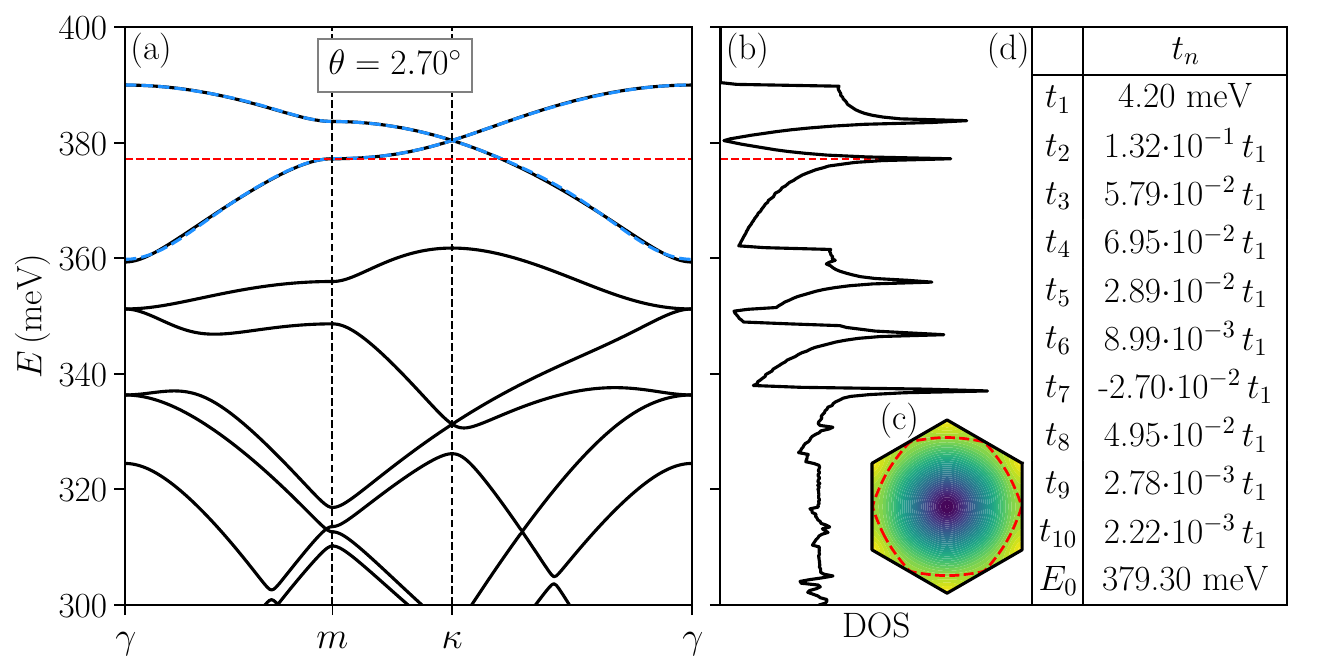}
    \includegraphics[scale=0.39]{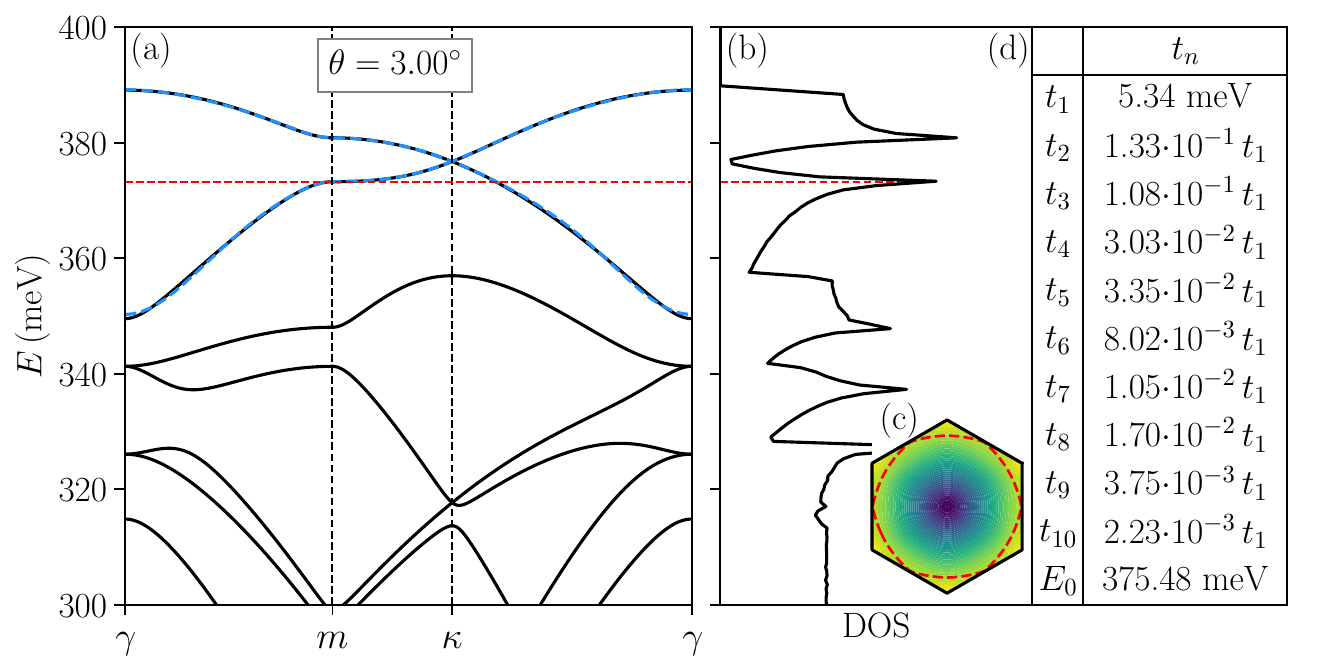}
    \includegraphics[scale=0.39]{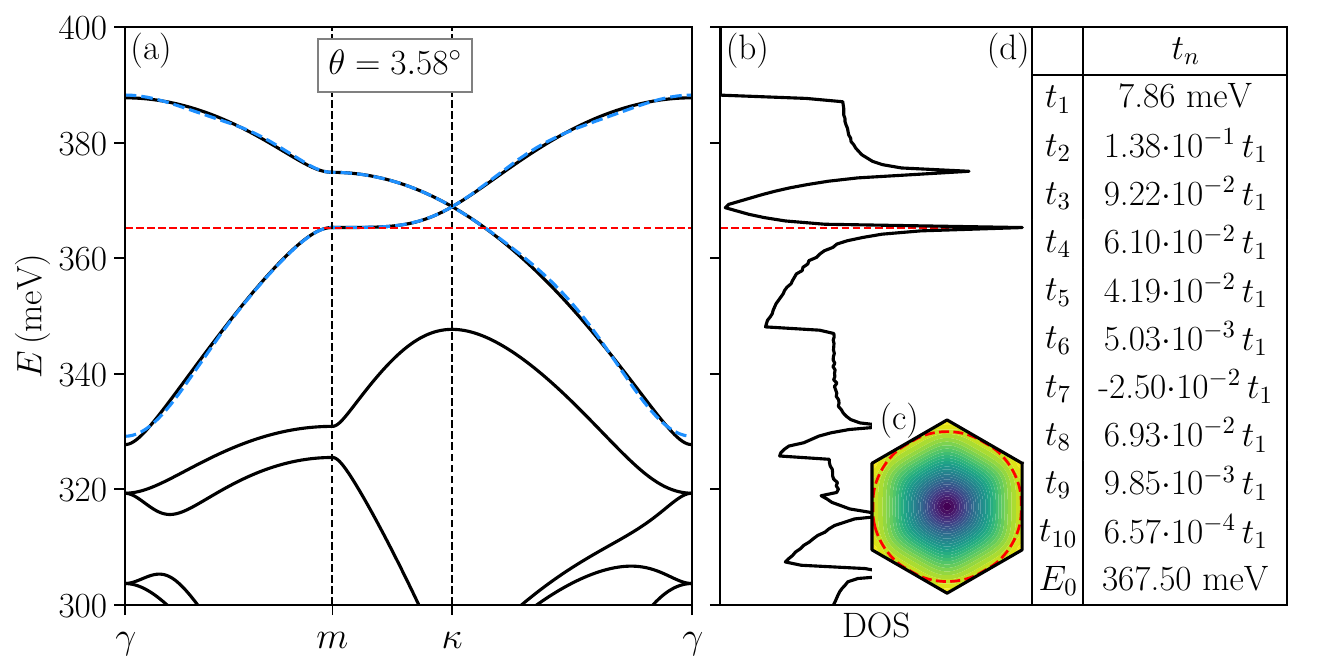}
    \includegraphics[scale=0.39]{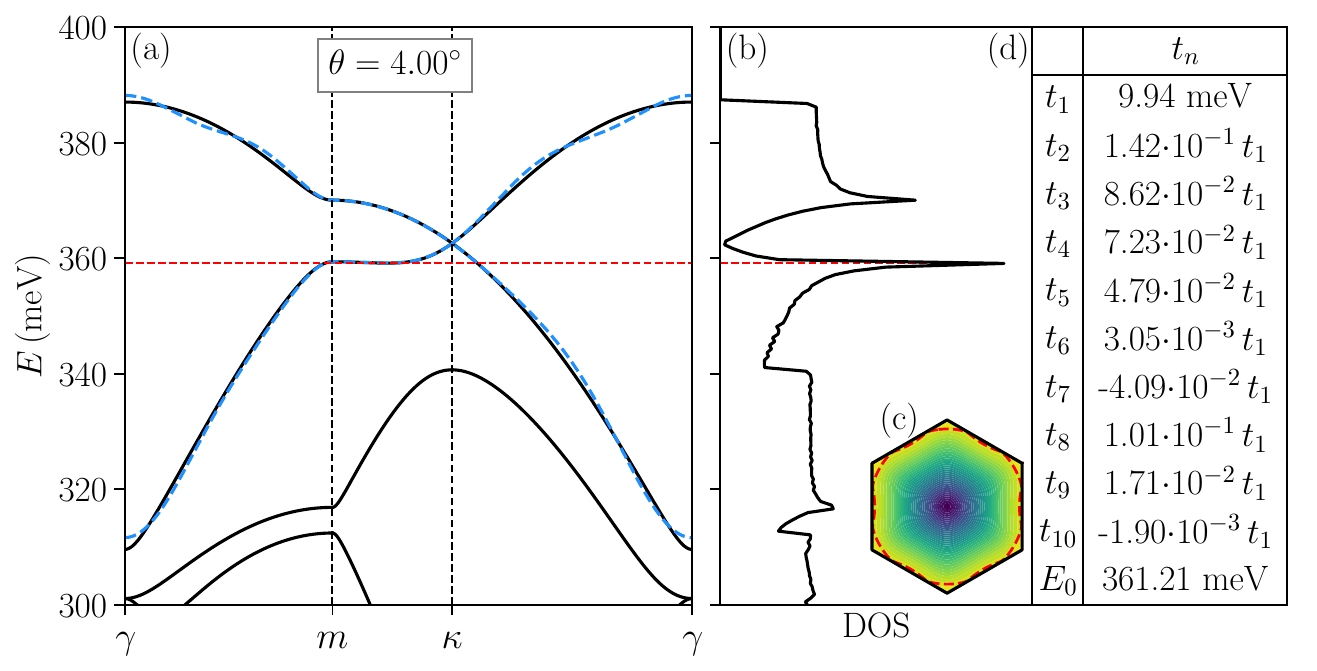}
    \caption{\textbf{Band structure from the continuum model for various twist angles.} (a) Band structure along the $\gamma$--$m$--$\kappa$--$\gamma$-path in the moir\'e BZ from the continuum model (black bands) and from the effective tight-binding model (blue bands), (b) corresponding density of states, and (c) Fermi surface at van Hove filling of the lower band of tdbWSe${}_2$ for twist angles $\theta \in \{2.0^\circ, 2.5^\circ, 2.7^\circ, 3.0^\circ, 3.58^\circ, 4.0^\circ\}$ from the continuum model with cutoff radius $4 G$. A magnification of these Fermi surfaces is shown in Fig.~\ref{fig:VHSsplitting}. Panel (d) shows the hopping amplitudes determined by fitting the honeycomb lattice tight-binding model defined in Eq.~\eqref{eq:tbhamil} to the two topmost bands of the continuum model.}
    \label{fig:bandstruc2}
\end{figure*}

\section{Band structure and twist angle}\label{sec:cont} 

\subsection{Continuum model}

We describe the $\Gamma$-valley moir\'e bands of ABBA-stacked \tdbWSe\ employing a continuum model~\cite{doi:10.1073/pnas.2021826118,Pan:2023}. 
The single-particle Hamiltonian describing the $\Gamma$~valley states reads
\begin{align}\label{eq:contmodel}
    H_{\vec{k},\sigma}^\mathrm{cont}(\vec{r}) = \frac{-\hbar^2 k^2}{2 m^*}+ 
    \begin{pmatrix}
        \Delta_1(\vec{r}) & \Delta_{12}(\vec{r}) & 0 & 0 \\
        \Delta_{12}^*(\vec{r}) & \Delta_2(\vec{r}) & \Delta_{23}(\vec{r}) & 0 \\ 
        0 & \Delta_{23}^*(\vec{r}) & \Delta_3(\vec{r}) & \Delta_{34}(\vec{r}) \\
        0 & 0 & \Delta_{34}^*(\vec{r}) & \Delta_4(\vec{r})
    \end{pmatrix}\,,
\end{align}
where $\vec{k}$ is the wavevector and ${\sigma = \uparrow, \downarrow}$ the spin polarization. 
We introduced the intralayer potentials $\Delta_i$ with $i = 1,2,3,4$, and the interlayer tunnelings $\Delta_{12}$, $\Delta_{23}$, and $\Delta_{34}$. For small twist angle between the second and third layers, the moir\'e reciprocal lattice vectors $\vec{G}_i$ appear in $\Delta_{2}$, $\Delta_3$, $\Delta_{23}$ as
\begin{align}
    \Delta_l(\vec{r}) &= V_2^{(0)} + 2 V_2^{(1)} \sum_{i=1,3,5} \cos(\vec{G}_i \cdot \vec{r} +(-1)^l \phi),\\
    \Delta_{23}(\vec{r}) &= V_{23}^{(0)} + 2 V_{23}^{(1)} \sum_{i=1,3,5} \cos(\vec{G}_i \cdot \vec{r})\,,
\end{align}
with $l\in\{2,3\}$.The potentials of the other layers and the tunneling between them are taken to be constant,
\begin{align}
    \Delta_1(\vec{r}) = \Delta_4(\vec{r}) = V_1,\quad\quad
    \Delta_{12}(\vec{r}) = \Delta_{34}(\vec{r}) = V_{12}.
\end{align}
The six first-shell moir\'e reciprocal lattice vectors are given by
\begin{equation}
    \vec{G}_i = \frac{4 \pi}{\sqrt{3} a_M} \left(\cos(\frac{(i-1) \pi}{3}), \sin(\frac{(i-1)\pi}{3})\right)^T
\end{equation} 
with $i = 1,2,...,6$ and moir\'e lattice constant $a_M$.

For small twist angles $a_M \approx a_0/\theta$, with WSe${}_2$ monolayer lattice constant $a_0 = 3.28\,\text{\AA}$. 
The model defined in Eq.~\eqref{eq:contmodel} features $C_{2y}$, $C_{3z}$, time-reversal, and $\mathrm{SU}(2)$ spin rotational symmetry due to negligible spin-orbit coupling near the $\Gamma$~valley~\cite{Pan:2023,Fang:2015}. It also has translational symmetry for any moir\'e lattice vector $\vec{R}$, i.e., $H_{\vec{k},\sigma}^\mathrm{cont}(\vec{r}+\vec{R}) = H_{\vec{k},\sigma}^\mathrm{cont}(\vec{r})$.

The following model parameters have been obtained by band-structure fits to untwisted ABBA double bilayer WSe${}_2$: $(V_1,V_2^{(0)}, V_2^{(1)})=(200,-159,-8)\,\mathrm{meV}$, $\phi=-0.17$, and $(V_{12},V_{23}^{(0)},V_{23}^{(1)})=(184,356,-9)\,\mathrm{meV}$~\cite{Pan:2023}. For the effective mass, we use the latest experimental results of Ref.~\cite{ma2024relativistic}, which finds $m^\ast=1.0\,m_\mathrm{e}$ with $m_\mathrm{e}$ being the bare electron mass.

Due to the translational symmetry, the single-particle Hamiltonian in Eq.~\eqref{eq:contmodel} can be directly diagonalized by a discrete Fourier transformation. We take into account all Fourier modes within a radius $4G = 16 \pi/\sqrt{3} a_M$, where $G=|\vec{G}_i|$, and we have explicitly verified that the inclusion of more modes leads to negligible quantitative improvement. 

\subsection{Angle-dependent band structure}

In Fig.~\ref{fig:bandstruc2}, we show the resulting continuum-model band structures along the $\gamma$--$m$--$\kappa$--$\gamma$-path in the Brillouin zone~(BZ) (panel (a)) and the corresponding DOS (panel (b)) for a series of twist angles ${\theta \in \{2.00^\circ, 2.50^\circ,2.70^\circ,3.00^\circ,3.58^\circ,4.00^\circ\}}$. 
The two topmost bands mimic the two-orbital band structure of a tight-binding model on the honeycomb lattice and the two concomitant VHS can be easily identified in the DOS.
For small angles, they appear for hole doping density near $\nu \approx 3/2$ and $\nu \approx 5/2$~\cite{ma2024relativistic}, in units of the moir\'e density, and they emerge from the energy dispersion at the $m,m',m''$ points. 
This coincides well with the prediction of the nearest-neighbor tight-binding model on the honeycomb lattice.

For all twist angles shown, we furthermore find that the band structure features a linear band crossing at the $\pm \kappa$~points at half filling ($\nu=2$). 
There the DOS vanishes linearly and the low-energy excitations can be described in terms of Dirac fermions.
We further observe that bandwidth and Fermi velocity at the Dirac points systematically decrease with decreasing twist angle, which goes in line with the fact that the moir\'e BZ decreases with decreasing twist angle, since  $1/a_M \propto \theta$.

\subsection{Tuning from ordinary to high-order van Hove singularity}

Topology and geometry of  electronic band structures are key to understanding the correlated states that emerge in the presence of interactions.
In two dimensions, a special role is played by the saddle points of an energy band around which the dispersion can be quadratically expanded as $\delta k_1^2-\delta k_2^2$. 
Here, $\delta k_{1,2}$ are the deviations from the position of the saddle point in wavevector space in two orthogonal directions. 
Saddle points are guaranteed to exist in two dimensions and generically lead to a VHS where the DOS is logarithmically divergent~\cite{PhysRev.89.1189}.
Tuning the Fermi level across the VHS, the Fermi-surface undergoes a topological (Lifshitz) transition~\cite{Lifshitz1960}.

In exceptional cases, known as high-order critical points, saddle-points cannot be described at leading order by a quadratic  expansion, but need higher-order terms. 
This leads to power-law divergent DOS, i.e., a high-order van Hove Singularity~(HOVHS).
HOVHS close to the Fermi level characteristically affect the transport and thermodynamic properties of the system~\cite{yuan2019magic,kerelsky2019maximized}, see Ref.~\cite{Classen_2025_flat_HOVH} for a review.
Moreover, the high-density of states suggests a strong susceptibility of such a system towards the formation and competition of interaction-induced correlated states~\cite{PhysRevB.95.035137,PhysRevResearch.1.033206,Classen:2020,PhysRevB.102.245122,PhysRevB.107.045122,PhysRevResearch.5.L012034,PhysRevResearch.5.L042006,PhysRevB.109.155118,beck2025kekul}.
%
\begin{figure}[t!]
    \centering
    \includegraphics[width=0.8\columnwidth]{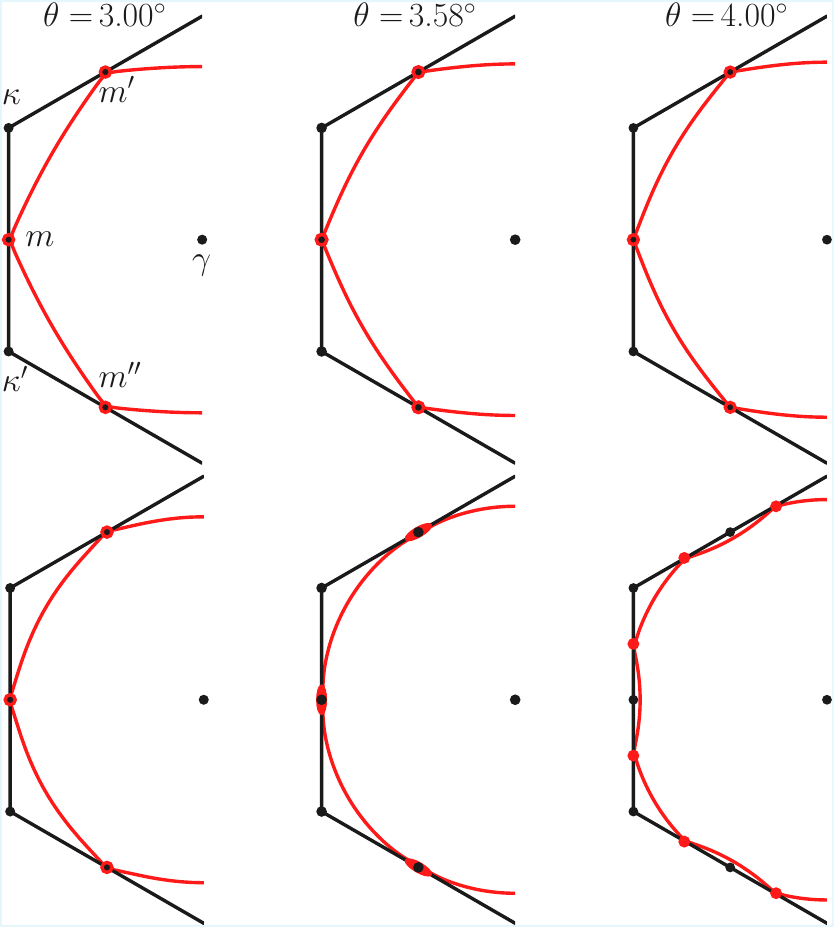}
    \caption{\textbf{Evolution of Fermi surface at van Hove fillings.} We show the evolution of the Fermi surface at the Van-Hove filling of the topmost (second-to-topmost) moir\'e band in the top (bottom) row for angles ${\theta \in \{3.00^\circ,3.58^\circ,4.00^\circ\}}$.
    In the lower band, each saddle point splits up into two, when crossing the angle $\theta_c\sim 3.58^\circ$ from below. Right at $\theta_c$ there is a degeneracy, leading to a high-order saddle point.
    }
    \label{fig:VHSsplitting}
\end{figure}
%
Reaching a HOVHS typically requires fine-tuning of system parameters, which is often difficult to achieve in conventional solids.
On the other hand, the high degree of tunability of moir\'e systems has opened possibilities to study HOVHS more precisely, e.g., for the case of twisted bilayer graphene~\cite{yuan2019magic}.
Here, we show that \tdbWSe\ is a prime platform for twist-angle tuning of a HOVHS.

To that end, consider the VHS in the moir\'e BZ of the two topmost moir\'e bands starting from small angles around $\sim 2^\circ$.
We find that the VHS at the $m$ points of the topmost moir\'e band survives for all twist angles up to $4^\circ$ and the Fermi surface at van Hove fillings only deviates mildly from being perfectly nested, see the top row in Fig.~\ref{fig:VHSsplitting}.
In contrast, in the second-to-highest moir\'e band,  the Fermi surface at van Hove filling is deformed away from approximate nesting at $\sim 2^\circ$ and becomes rounder for larger angles. 
The saddle points stay at the $m$~points at first, but for $\theta_\mathrm{HOVHS} \approx 3.58^\circ$, the Fermi surface at van Hove filling snuggles against the edge of the moir\'e BZ, see the middle panel in the second row of Fig.~\ref{fig:VHSsplitting}.
For even larger angles, the saddle points at the $m$~points split into two each, and move towards  $\kappa,\kappa'$ along the $m$-$\kappa^{(\prime)}$ lines, see Fig.~\ref{fig:VHSsplitting}.

The splitting of the saddle point into two indicates the emergence of a high-order critical point~\cite{Classen_2025_flat_HOVH}. 
This can be corroborated by a Taylor series expansion of the energy band at the $m$~points, which generically starts with contributions quadratic in the deviations from the $m$~points.
We observe that at the transition where the saddle point splits into two, i.e., at $\theta_\mathrm{HOVHS} \approx 3.58^\circ$, the leading terms of the expansion -- after appropriate rescaling -- are given by
$\epsilon_{\mathrm{HOVHS}}(m+ \delta \vec{k})\sim \delta k_1^4 - \delta k_2^2\,,
$
where we dropped an irrelevant perturbation $\sim \delta k_1^2 \delta k_2^2$.
The corresponding DOS induced by this high-order saddle-point dispersion diverges as $|\delta \epsilon|^{-1/4}$, which is robust with respect to higher order corrections in the expansion~\cite{Classen_2025_flat_HOVH}.
The precise angle of the real material can be expected to be somewhat different from our prediction of  $\theta_\mathrm{HOVHS} \approx 3.58^\circ$ due to the approximations that go into the continuum model.
The splitting of saddle points and the concomitant HOVHS can, however, be expected to be a robust property of the model.

The presence of such HOVHS can be experimentally confirmed by scanning tunneling spectroscopy, where the tunneling conductance shows a peak when crossing a VHS:
at a HOVHS, the conductance peak follows the power-law scaling from the DOS $\sim |\delta \epsilon|^{-1/4}$~\cite{yuan2019magic}.
Our band-structure calculations suggest that this behavior can be observed in \tdbWSe.

Interestingly, real graphene can also be doped to van Hove filling by employing intercalation techniques~\cite{PhysRevLett.104.136803,PhysRevB.100.121407,PhysRevLett.125.176403}. 
In this system a band flattening around the $M$~points, similar to the one observed here, has been found, which may turn the  conventional VHS into a HOVHS, too.
This strengthens the narrative of tdbWSe${}_2$ being a simulator for strongly-correlated graphene even beyond Dirac filling.

\section{Effective Hubbard model}
\label{sec:effhubbard}

To study strongly-correlated phases of \tdbWSe, we construct an effective Hubbard model.
To that end, we first map the two valence bands to an effective tight-binding model on the moir\'e honeycomb lattice corresponding to the moir\'e BZ from the continuum model, cf. Sec.~\ref{sec:cont}.

\subsection{Honeycomb-lattice tight-binding part}\label{sec:tbparam}

We write the tight-binding part of the Hamiltonian on the moir\'e honeycomb lattice as
\begin{align} \label{eq:tbhamil}
    H_0 = -\sum_{n = 1}^{N_h} \sum_{i,j,\sigma} t^{(n)}_{ij}(\theta) c_{i\sigma}^\dagger c_{j\sigma}\,.
\end{align}
Here, $c^\dagger_{i,\sigma}$ creates a fermion with spin polarization~$\sigma$ on site~$i$ and the angle-dependent hoppings $t^{(n)}_{ij}(\theta)$ are defined as
\begin{align}
    t^{(n)}_{ij}(\theta) = \begin{cases}
       \ t_n(\theta), & \text{if} \quad d_{n-1} < |\vec{R}_i - \vec{R}_j| < d_{n + 1}, \\
       \ 0, &\text{else}\,.
    \end{cases}
\end{align}
In the above, $\vec{R}_i$ denote the moir\'e lattice vectors, $d_n$ is the distance of a lattice site to its $n$-th nearest neighbor, and $N_h$ denotes the number of hoppings taken into account.

We determine the hopping amplitudes $t_n$ by fitting the tight-binding band structure to the band structure obtained with the continuum model.
In the following, we use $N_h = 10$ hopping amplitudes, which provides a good balance between the average relative deviation from the continuum model and the faithful reproduction of the analytic structure of the bands, see App.~\ref{app:fitting} for details and quantitative measures for the quality of our fitting procedure.
To match the band energies from the continuum model, we introduce a constant energy shift~$E_0$.
We show the resulting tight-binding parameters, the energy shift~$E_0$, and the corresponding bands obtained from the fitted tight-binding model as the blue dashed lines in Fig.~\ref{fig:bandstruc2}.
In our data files~\cite{ourdata2025}, we also provide the tight-binding parameters $t_n(\theta)$ for $\theta\in [1.0^\circ,4.0^\circ]$ in steps of $0.02^\circ$. 
We note that in the fitted tight-binding model, the high-order van Hove singularity appears at a slightly shifted value of~$\theta_\text{HOVHS} = 3.51^\circ$.

\subsection{Hubbard interaction}

We take into account electron-electron interactions by including an on-site Hubbard repulsion given by
\begin{equation}\label{eq:hubbardU}
    H_\mathrm{int} = U(\theta) \sum_i c_{i\uparrow}^\dagger c_{i\downarrow}^\dagger c_{i\downarrow} c_{i\uparrow}.
\end{equation}
In order to estimate the angle-dependent strength of the on-site Hubbard interaction $U(\theta)$, we employ an approximation scheme put forward in Ref.~\cite{Wu:2018}. 
To that end, first note that the maxima of the moir\'e potential associated with the four layers in Eq.~\eqref{eq:contmodel} coincide to a good approximation (due to the small phase shift of $\phi = -0.17$) and are located on the underlying triangular lattice. 
Wannier orbitals will then be localized around these maxima, and the on-site interaction can then be estimated by ${U \sim e^2 / 4 \pi \epsilon a_W}$, where $a_W$ is the spread of the Wannier function. 
In the vicinity of the maxima, the potential can be approximated quadratically, yielding $a_W \propto \sqrt{a_M}$~\cite{Wu:2018}. 
Hence, $a_W \propto 1/\sqrt{\theta}$, and the on-site Coulomb energy scales as~$U \propto \sqrt{\theta}$. 

The proportionality constant can be fixed by providing the interaction at a reference angle $\theta_0$, yielding 
\begin{align}\label{eq:hubbardUtheta}
U(\theta) = U(\theta_0) \sqrt{\theta/\theta_0}\,.
\end{align}
Here, we choose $\theta_0 = \theta_c = 2.7^\circ$, where the experiment of Ref.~\cite{ma2024relativistic} finds a transition from a Dirac semimetal to an insulator.
We note that the hopping amplitudes decrease faster than $\propto \sqrt{\theta}$ towards smaller angles, i.e., the dimensionless ratio $U(\theta)/t_1(\theta)$ increases towards small $\theta$. Hence, the effect of interactions is systematically enhanced when $\theta$ decreases.

\begin{figure}[t!]
    \centering
    \includegraphics[width=0.9\columnwidth]{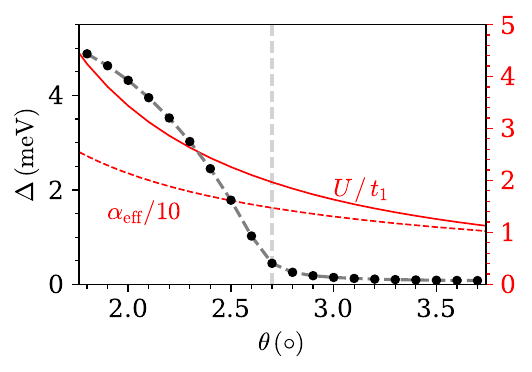}
    \caption{\textbf{Hubbard interaction and relativistic Mott transition} from Hartree-Fock mean-field approximation. 
    We show the antiferromagnetic gap $\Delta = 2 U m$ for different twist angles $\theta$ determined with the effective Hubbard model described in Sec.~\ref{sec:effhubbard} and $\mathcal{N} = 24 \times 24$ unit cells (black dots), and the angle-dependent quantities $U(\theta)/t_1(\theta)$ (red curve) and $\alpha_\text{eff}$ (red dashed curve). We rescaled the effective fine-structure constant by a factor of 1/10 for better visibility. At $\theta_c = 2.7^\circ$, a transition from an antiferromagnet ($\theta < \theta_c$) to a Dirac semimetal ($\theta > \theta_c$) occurs. We note that the small but finite gap at $\theta_c$ is a finite-size effect, as we determined an estimate for~$U_c$ in the thermodynamic limit from the correlation ratio in Eq.~\eqref{eq:corrrat}.
    }
    \label{fig:Mott}
\end{figure}

The half-filled honeycomb lattice Hubbard model requires a critical interaction strength $U_c$ for the formation of an insulating state~\cite{Sorella_1992,Meng_2010,Sorella_2012,PhysRevB.86.045105,PhysRevX.3.031010,PhysRevB.89.205128,PhysRevB.101.125103}.
This suggests to choose $U(\theta_0)$ such that the effective Hubbard model coincides with $U_c$.
We do this within a Hartree-Fock mean-field approximation and show our result for $U(\theta)/t_1(\theta)$ in Fig.~\ref{fig:Mott}. 
Further details of the Hartree-Fock study are discussed in the next section.
In~\cite{ourdata2025}, we provide our estimates for the Hubbard interaction $U(\theta)$ for $\theta\in [1.0^\circ,4.0^\circ]$ in steps of $0.02^\circ$ along with the tight-binding parameters. 

An alternative measure for the strength of the Coulomb interaction is the fine-structure constant where the speed of light is replaced with the angle-dependent Fermi velocity. This ``effective'' fine-structure constant~$\alpha_\text{eff}$  is directly related to the Wigner-Seitz radius~\cite{RevModPhys.83.407}, and can also be taken as an estimate for~$U/t_1$. For reference, we show~$\alpha_\text{eff}$ in Fig.~\ref{fig:Mott}, where we can see that it is larger than our Hartree-Fock-based estimate for~$U/t_1$, but roughly in the same order of magnitude.

\section{Hartree-Fock mean-field calculations} 

\subsection{Hartree-Fock Method}

We consider the effective  Hubbard model with~$\mathcal{N}$ unit cells, i.e., $N = 2\mathcal{N}$ sites, and periodic boundary conditions. The Hamiltonian is $H = H_0 + H_{\mathrm{int}}$ with hoppings $t_{ij}^{(n)}(\theta)$, on-site repulsion $U(\theta)$, and $U(\theta_0)$ to be determined. 
We calculate at fixed temperature $T$ and filling $n$. 
We introduce a mean-field decoupling of the four-fermion terms by a suitable bilinear, yielding~\cite{Zaanen1989,Scholle2023hubbard,scholle2024mean}
\begin{align}
    \label{eq:HMFT} H_{\mathrm{int}}^{\mathrm{MF}} = &\sum_{j,\sigma} \Delta_{j\overline{\sigma}}n_{j\sigma} + \sum_{j} \left( \Delta_{j-} c_{j\uparrow}^\dagger c_{j\downarrow} + \Delta_{j+} c_{j\downarrow}^\dagger c_{j\uparrow} \right)\nonumber \\
    &- \frac{1}{U_0} \sum_j\left( \Delta_{j\uparrow}\Delta_{j\downarrow} - \Delta_{j-}\Delta_{j+} \right) \,,
\end{align}
with $\overline{\uparrow}=\ \downarrow$, $\overline{\downarrow}=\ \uparrow$, and $4N$ mean-field parameters ${\Delta_{j\sigma} = U_0 \langle n_{j\sigma}\rangle}$ and ${\Delta_{j+} = \Delta_{j-}^* = -U_0 \langle c_{j\uparrow}^\dagger c_{j\downarrow} \rangle}$. 

The full mean-field Hamiltonian is $H^{\mathrm{MF}} = H_0 + H_{\mathrm{int}}^{\mathrm{MF}}$. Since $H^{\mathrm{MF}}$ is bilinear, the eigenvalue problem reduces to the diagonalization of a $2N \times 2N$ dimensional single-particle Hamiltonian. For details on our Hartree-Fock scheme, see App.~\ref{app:hfmft}.

\subsection{Observables}

The solutions from the Hartree-Fock scheme are used to calculate observables. 
We parametrize the spin operator by 
$
\vec{S}_i = \frac{1}{2} \vec{c}_i^\dagger \sigma_i \vec{c}_i$ with $\vec{c}_i = (c_{i\uparrow}, c_{i\downarrow})^T
$
and Pauli matrices $\sigma_i$. 
The expectation value of $\vec{S}_i$ for a given configuration in terms of the mean-field parameters reads
\begin{align}
    \hspace{-0.2cm}\langle S_{j}^x \rangle\! = \! \frac{\Delta_{j+}\! +\! \Delta_{j-}}{-2U},\ \langle S_{j}^y \rangle \! = \!  \frac{\Delta_{j+} \! +\! \Delta_{j-}}{-2Ui},\
    \langle S_{j}^z \rangle \! = \frac{\Delta_{j\uparrow} \! -\! \Delta_{j\downarrow}}{2U_0}.
\end{align}
Any spin-ordered state induces a finite absolute magnetization,
\begin{align}\label{eq:mag}
    m = \frac{1}{N} \sum_{i=1}^N \sqrt{\langle S_{i}^x \rangle^2 + \langle S_{i}^y \rangle^2 + \langle S_{i}^z \rangle^2},
\end{align}
while a paramagnetic phase is uniquely classified by $m=0$ in the thermodynamic limit, or, equivalently, $\langle \vec{S}_j\rangle = 0$.
In the case of (perfect) ferromagnetism, $m$ coincides with the length of the magnetization $\vec{m}_\mathrm{FM} = \sum_i \langle \vec{S}_i \rangle /N$, while in the case of antiferromagnetism, it serves as an upper bound for the staggered magnetization $\vec{m}_\mathrm{AFM} = \sum_i (-1)^i \langle \vec{S}_i \rangle /N$. 

A magnetic state with $m > 0$ can be further classified by analyzing the modes $\vec{S}_{\vec{q}}$ appearing in the Fourier transform
\begin{equation}\label{eq:spinfourier}
    \langle \vec{S}_{i} \rangle = \langle \vec{S}_{j,\lambda} \rangle = \sum_{\vec{q}} \vec{S}_{\vec{q},\lambda} e^{i \vec{q} \cdot \vec{R}_j},
\end{equation}
where $\vec{q}$ runs over the BZ of the moire reciprocal lattice, and we replaced the site index~$i$ with a Bravais lattice site index~$j$ and sublattice index~$\lambda \in \{A,B\}$. 
We assume that $\langle \vec{S}_{j,\lambda} \rangle$ is independent of the sublattice, i.e., we set $\vec{S}_{\vec{q},\lambda} = \vec{S}_{\vec{q}}$. Different ordered states manifest themselves in different dominant modes $\vec{S}_{\vec{q}}$ appearing in the Fourier transform in Eq.~\eqref{eq:spinfourier}.

A paramagnetic state with $\langle \vec{S}_i \rangle = 0$ has only vanishing Fourier modes, i.e., $\vec{S}_q = 0$. 
Collinear magnetism, on the other hand, is characterized by $\vec{S}_{\vec{q} = 0} > 0$, while $\vec{S}_{\vec{q} \neq 0} = 0$. 
Similar to Refs.~\cite{Sachdev2019stripe,Millis2022,Scholle2023hubbard,scholle2024mean}, we further distinguish between various kinds of stripe orders. 
First note that the real valued-ness of $\langle \vec{S}_i \rangle$ implies $\vec{S}_{\vec{q}} = \vec{S}^*_{-\vec{q}}$. 
Hence the number of dominant modes is even, and only half of them are independent. 
Different types of stripe orders differ in the number $N_\text{dom}$ of independent dominant modes $\vec{S}_{\vec{Q}_k}$ as well as the geometry between them.
Here, we distinguish between single-mode stripe order, two orthogonal stripes, and three modes with the same amplitude, where $N_\text{dom}=1,2,3$, respectively. 
The latter can be further divided into phases where the three modes are orthogonal, where they are collinear, and where none of the two applies. If a magnetic state does not fall in any of the phases above, we classify it as ``other'' order. Note that in a finite system,  $m$ can be finite but small even in the paramagnetic phase, and the free energy is always analytical. Hence, for the classification of a state as paramagnetic we choose a threshold, see App.~\ref{app:class}.

For a more precise classification of phases in terms of Fourier modes $\vec{S}_{\vec{q}}$ and their geometry, we refer to Refs.~\cite{Scholle2023hubbard,scholle2024mean} and App.~\ref{app:hfmft}.
We specify further physical properties of the dominant magnetic states that we find in our Hartree-Fock mean-field approach to \tdbWSe, below.

\section{Relativistic Mott transition}\label{sec:relmott}

In Ref.~\cite{ma2024relativistic}, a transition from a semimetal to an insulator has been observed in \tdbWSe\ at $\theta_c = 2.7^\circ$. 
We use this to fix the interaction strength of our effective Hubbard model, cf. Eq.~\eqref{eq:hubbardU}, within the Hartree-Fock approach.
The corresponding hopping parameters are already listed in Fig.~\ref{fig:bandstruc2}.

To proceed, we first determine the value $U_c$ as a function of~$\theta$, where a transition from a semimetallic to an insulator occurs at Dirac filling. 
We then set $U(\theta_0) = U_c(\theta_0)$, where $\theta_0 = \theta_c = 2.7^\circ$, which  fixes the Hubbard interaction for all angles $\theta$, cf. Eq.~\eqref{eq:hubbardUtheta}. 
Compatibility with the experimentally observed transition is then ensured by construction.
Our half-filled Hubbard model transitions into a N\'eel antiferromagnet for $U>U_c$.

In order to reliably determine the critical interaction strength $U_c$ at which such a phase transition occurs in the thermodynamic limit, we consider the correlation ratio~\cite{Fisher_1967,PhysRevLett.115.157202,kennedy2025extended}
\begin{equation} \label{eq:corrrat}
    R(N,U)=1-\frac{S_{\mathrm{AFM}}(\vec{q} + \delta\vec{q})}{S_{\mathrm{AFM}}(\vec{q})}\,.
\end{equation}
Here, the vector $\delta \vec{q}$ is a nearest-neighbor reciprocal lattice vector and $S_{\mathrm{AFM}}=S_{\mathrm{AFM}}(\vec{q})$ is the static antiferromagnetic structure factor, reading
\begin{align}
    S_{\mathrm{AFM}}=\frac{1}{N}\!\sum_{i,j}\! e^{-i\vec{q}\cdot (\vec{R_i}\!-\!\vec{R_j})} \big\langle (S_{i,A}^z\! -\! S_{i,B}^z)(S_{j,A}^z \!-\! S_{j,B}^z) \big\rangle.
\end{align}
In the above, $S_{i,\lambda}^z$ denotes the $z$~component of the spin operator at Bravais lattice site $i$ and sublattice $\lambda \in \{A,B\}$. The system-size dependent correlation ratio $R$ is defined such that (1)~it vanishes in the semimetallic phase, (2)~it is non-vanishing in the antiferromagnetic phase and approaches one in the thermodynamic limit, and (3)~it is system-size independent precisely at the transition, i.e., close to $U_c$, it follows the functional form $R(N,U) = \mathcal{F}[(U-U_c)N^{1/2\nu}]$~\cite{PhysRevLett.115.157202}.
Hence, the curves of $R(N,U)$ measured for various finite system sizes intersect at a single point as a function of $U$, and the intersection point coincides with $U_c$ in the thermodynamic limit $N \to \infty$.
Based on this procedure, we extract $U_c/t_1 \approx 1.974$ for the angle $\theta_c = 2.7^\circ$ where the experimental data locates the relativistic Mott transition. 
We show further data for $t_1(\theta)$ as obtained from our tight-binding fitting procedure, $U_c(\theta)$ determined by the correlation ratio, and~$U(\theta)$ in App.~\ref{app:hfmft}.

In Fig.~\ref{fig:Mott}, we present the resulting Hubbard interaction~$U(\theta)$ in units of the hopping $t_1(\theta)$ as the red curve, along with the magnitude of the antiferromagnetic gap ${\Delta = 2 U m}$ as a function of the twist angle for $\mathcal{N} = 24 \times 24$ unit cells.
We note that for the half-filled honeycomb-lattice Hubbard model with nearest-neighbor hopping only, the mean-field approach overestimates the tendency towards the formation of order. 
It already finds the transition at $U_c^{\mathrm{MFT}}\approx 2.23 t_1$ while quantum Monte Carlo simulations find $U_c^{\mathrm{QMC}}\approx 3.8 t_1$, see, e.g., Refs.~\cite{Sorella_2012,PhysRevX.3.031010,PhysRevB.101.125103}.
Hence, our estimate for $U(\theta)/t_1(\theta)$ is expected to be smaller than the ``real'' value.
Nevertheless, the order of magnitude of our result for the antiferromagnetic gap coincides with the experimentally determined values~\cite{ma2024relativistic}.
The antiferromagnetic insulator that we find for small angles suggests that the corresponding quantum phase transition  belongs to the Gross-Neveu-Heisenberg universality class~\cite{Rosenstein:1993zf,PhysRevB.80.075432,Janssen2014afm,PhysRevB.97.075129,PhysRevD.96.096010,Gracey2018heisenberg,Ladovrechis2023heisenberg,Biedermann:2025dma,PhysRevX.3.031010,PhysRevX.6.011029,PhysRevB.102.235105,PhysRevB.98.235129,PhysRevB.99.205434,PhysRevB.102.245105,PhysRevB.104.155142}.

\section{Hartree-Fock phase diagram}\label{sec:hfphasediagram} 

\begin{figure*}[t!]
    \centering
    \includegraphics[width=0.85\textwidth]{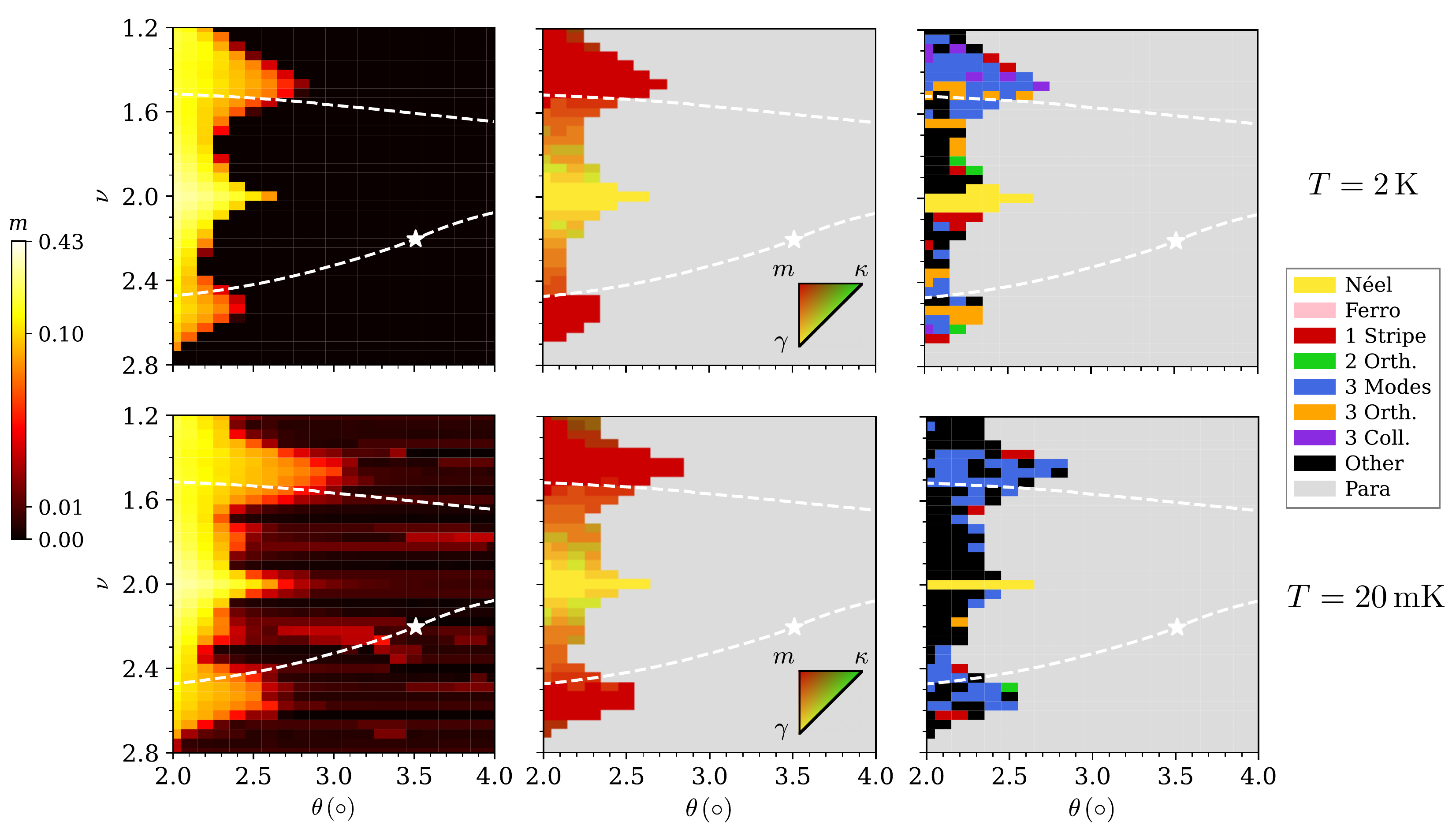}
    \caption{\textbf{Absolute magnetization, momentum transfer, and phase diagram in the $n$-$\theta$ -- plane for $T = 20\,\mathrm{mK}$ and $T = 2\,\mathrm{K}$.} Absolute magnetization on the finite lattice (left panels), momentum transfer (middle panels), and phase diagram (right panels) in the filling--twist-angle plane at $T = 20\,\mathrm{mK}$ (bottom panels) and $T = 2\,\mathrm{K}$ (top panels) from Hartree-Fock calculation for a lattice with $18 \times 18$ unit cells. The dashed white lines mark the van Hove fillings of the upper and lower band, and the star marks the position of the higher-order van Hove singularity at $\theta = 3.51^\circ$. 
    We note that the paramagnetic regions (grey) in the middle and right panels have been determined by using a threshold to consider finite-size effects, see App.~\ref{app:class}. 
    In the legend, ``1 Stripe'' refers to single-mode stripe order, ``2 Orth.'' to two orthogonal stripes, ``3 Modes'' to three modes with the same amplitude, ``3 Orth.'' to orthogonal stripes, and ``3 Coll.'' to collinear stripes. The latter two are special cases of the ``3 Modes'' order, see the main text and App.~\ref{app:class} for more details.}
    \label{fig:phasediagram}
\end{figure*}

The data on the relativistic Mott transition at half-filling from Sec.~\ref{sec:relmott} completes the determination of parameters of the effective Hubbard model for \tdbWSe, yielding the angle-dependent interaction strength $U(\theta)$.
Together with the tight-binding parameters $t_n(\theta)$ found in Sec.~\ref{sec:tbparam}, this allows us to calculate the Hartree-Fock mean-field phase diagram at and away from half-filling, and at different temperatures.

We map out the full filling--twist-angle mean-field phase diagram for the experimentally accessible temperatures ${T = 20\,\text{mK}}$ and ${T = 2\,\text{K}}$.
We note that such Hartree-Fock mean-field calculations typically overestimate the tendency towards the formation of order and do not fulfill the Mermin-Wagner theorem. 
Hence, the magnetic orders that we identify, here, at finite~$T$, should rather be considered as representing the dominant correlations and may guide more advanced many-body approaches.

In Fig.~\ref{fig:phasediagram}, we show our results for ${T = 20\,\text{mK}}$ (bottom panels) and ${T = 2\,\text{K}}$ (top panels). More precisely, we provide data for the following:
\begin{enumerate}\setlength\itemsep{0em}
    \item The absolute magnetization, cf. Eq.~\eqref{eq:mag}, 
    \item The transfer momentum of the dominant spin mode in the Fourier expansion in Eq.~\eqref{eq:spinfourier},
    \item The classification of emergent magnetic orders.
\end{enumerate}

At half-filling for $\theta < 2.7^\circ$ at $T=20\,\text{mK}$, we robustly find antiferromagnetic N\'eel order with momentum-transfer $\vec{q} = 0$, as discussed before. 
Moving away from half-filling, we find two additional domains with increased magnetization around the two van Hove fillings of the lower and upper bands that persist to twist-angles around $2.7^\circ$. 
In both regions, the dominant Fourier modes have wavevectors near the $m, m'$, and $m''$ points of the moir\'e BZ.
This indicates that the magnetic order is not of collinear nature, see below. 
Instead, our Hartree-Fock calculations suggest that the order is most likely a stripe pattern with three dominant modes of same amplitude. 

To obtain our results, we have iterated the self-consistency loop of the Hartree-Fock scheme until a precision of $1\cdot 10^{-9}$ or a number of $10,000$ steps are reached, and we have repeated this procedure three times starting from different random initial conditions for each point in parameter space $\{\theta,n,T\}$.
As it turns out, the free energy landscape consists of various local minima that are very close to each other, e.g., at some points, their free energies differ by less than $\mathcal{O}(10^{-4})$ when going away from half filling.
Interestingly, we find that the absolute magnetization and the dominant transfer momentum are almost identical in the different local minima, i.e., these two quantities are robust and characteristic features of the emergent magnetic order. 
In contrast, each of these minima correspond to different stripe patterns with three dominant modes,  and the small energy difference is caused by a competition between the respective order parameters. 
Hence, due to the tiny differences in their free energies, the precise determination of the energetically favored magnetic order is not unambiguous and within our numerical implementation, we still find a non-negligible dependence on the initial conditions.

We can compare our results to the paradigmatic honeycomb-lattice Hubbard model with nearest-neighbor hopping, only.
A detailed Hartree-Fock mean-field study was recently put forward in Ref.~\cite{scholle2024mean} as a benchmark for the investigation of Bernal bilayer graphene.
We find our implementation to be consistent with the results presented there.
In particular, for smaller angles $\theta \sim 2^\circ$, where the hopping ratios $t_{n>1}/t_1$ are systematically smaller than for larger angles and the $U/t_1$ is between 3 and 4, we find that the magnetic orders agree very well with the ones in Ref.~\cite{scholle2024mean}.
Minor differences can be explained by the tiny deviations of the free energies of different magnetic states that may induce a preference of one at the expense of another if the hoppings are slightly different.

In summary, our Hartree-Fock calculations provide robust statements about the presence or absence of order and the accompanied most-dominant momentum transfer. 
However, the predictions of the precise nature of the order have to be taken with caution. 
We interpret the close competition of various different magnetic orders in the phase diagram of the \tdbWSe\ model as a characteristic feature of the material.

In the following, we discuss a series of noticeable aspects of the phase diagram.
First, we note that the experimental data~\cite{ma2024relativistic} considers the temperature dependence of the longitudinal resistivity to identify metallic or insulating behavior.
For half filling, a transition from one to the other appears at $2.7^\circ$ at low temperatures, which is -- by construction of our Hubbard interaction -- in agreement with our theoretical calculations, cf. Fig.~\ref{fig:Mott}.
At a twist of $3.0^\circ$, our data suggests that no magnetic order appears over the whole range of fillings calculated, neither at $2\,\mathrm{K}$ nor at $20\,\mathrm{mK}$.
Notably, this includes the van Hove fillings in both honeycomb-lattice bands where, due to the high DOS, interaction effects are enhanced and may drive the formation of order.
The absence of order is also in agreement with the experimental finding of metallic behavior in the range of filling factors between $\nu\approx 1.4$ and $\nu\approx 2.2$ for temperatures down to $20\,\mathrm{mK}$~\cite{ma2024relativistic}.

At $2.5^\circ$, where the ratio $U/t_1 \approx 2.1$ is higher, we find that, in addition to the N\'eel state at half filling, magnetic order occurs in a narrow range of fillings slightly above (below) the upper (lower) van Hove filling at low temperatures of $20\,\mathrm{mK}$.
While details depend on the precise filling and temperature, we generally observe that the magnetic order is characterized by three modes with the same amplitude and wavevectors at or near the $m,m',m''$ points.
Most of the states falling into that category gap out the entire Fermi surface and are hence insulating~\cite{Li_2012,PhysRevLett.101.156402}, with exception of the collinear stripes~\cite{PhysRevLett.108.227204}, which we comment on further, below.
Away from half filling, the experiment, instead, exclusively finds metallic behavior down to temperatures of $2\,\mathrm{K}$ in the whole range of explored fillings, including the van Hove filings~\cite{ma2024relativistic}.
We note that the angle $2.5^\circ$ is just slightly below the threshold where order appears in the first place and our calculations also find that at $2\,\mathrm{K}$ the magnetic order near the lower van Hove filling disappears, which suggests that it is quite fragile.
Hence, in the experiment, where also a certain extent of disorder is present, the possible order might be suppressed.
Also, it is known that mean-field calculations miss important fluctuation effects and, e.g., do not respect the Mermin-Wagner theorem. 

\begin{figure}[t!]
    \centering
    \includegraphics[width=0.9\columnwidth]{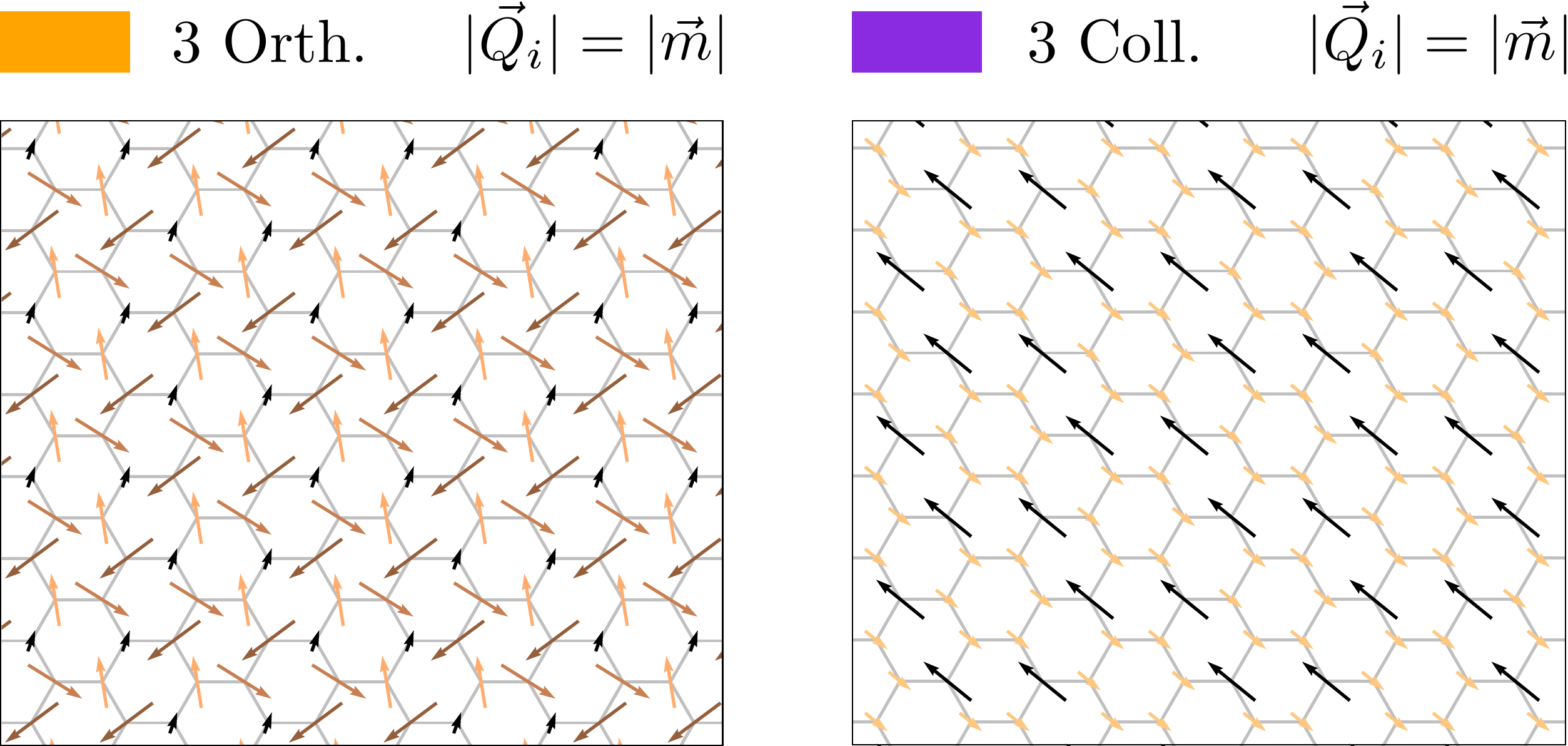}
    \caption{\textbf{Three orthogonal stripes and three collinear stripes} spin-ordering patterns at commensurate wavevector $\vec{Q_k}=|\vec{m}|$ on the real-space moir\'e honeycomb lattice.
    Left panel: This state corresponds to a non-coplanar SDW with non-zero spin chirality and was previously discussed in Refs.~\cite{Li_2012,PhysRevLett.101.156402}. We show the projection of the spins to the $x-z$-plane. The color of the arrow indicates the value of the $S_y$ component. Right panel: This state corresponds to a uniaxial SDW with only one spin branch being gapped out, i.e., it is a ``half metal'', which was described in Ref.~\cite{PhysRevLett.108.227204}.}
    \label{fig:mag}
\end{figure}

Based on our calculations, we predict, however, that magnetic order is stabilized for even smaller twist angles, where interactions become stronger.
An interesting candidate state near van Hove filling that we find within our set of magnetic orders is the one with three modes with the same amplitude where, additionally, the three amplitudes are orthogonal (``3 Orth.'', orange in Fig.~\ref{fig:phasediagram} and Fig.~\ref{fig:mag}).  In addition, the corresponding wavevectors are located at the $m,m',m''$ points.
This is a non-coplanar spin-density-wave~(SDW) state with non-zero spin chirality and quantized Hall conductance, cf. Refs.~\cite{Li_2012,PhysRevLett.101.156402}.
We note that this state appears at $2\,\mathrm{K}$ in our mean-field calculation, but transitions into another three-mode state at smaller temperatures.

An important robust feature of the magnetic order found near the van Hove fillings is its composition out of three modes with the same amplitude located at or very near the $m,m'$, and $m''$ points where the corresponding wavevectors are commensurate.
In addition to the previously discussed non-coplanar SDW, this further includes, e.g., the special case of the collinear stripe order (``3 Coll.'', purple in Fig.~\ref{fig:phasediagram}).
We show the spin-ordering pattern of this state on the moir\'e honeycomb lattice explicitly in the right panel Fig.~\ref{fig:mag}.
In Ref.~\cite{PhysRevLett.108.227204} the collinear stripe order with commensurate wavevectors at the $m,m',m''$ points is described as uniaxial SDW order and it corresponds to a ``half-metal'' state where only one spin branch is gapped out, while for the other one the original Fermi surface is preserved.
This leads to the possibility to electrically control the spin currents, which may be interesting for nanoscience applications~\cite{PhysRevLett.108.227204}.
Near van Hove filling, this state's free energy is very close to the other magnetic states with three same modes and it is therefore possible that it may be the dominant one in the real system.
Due to the half-metal nature of the uniaxial SDW, this would reconcile the experimental observation of metallic behavior in Ref.~\cite{ma2024relativistic} near the upper van Hove singularity down to $2\,\mathrm{K}$ with our data.

Finally, we discuss the HOVHS appearing in the lower honeycomb band at a twist of $\theta\sim 3.5^\circ$. 
There, the high-density of states suggests a strong susceptibility for the formation of interaction-induced order.
We have marked that point in the phase diagrams of Fig.~\ref{fig:phasediagram} with a star.
However, we do not find a prominent increase in the magnetization of our system at the HOVHS.
This can be explained by the small Hubbard interaction with $U(\theta_\text{HOVHS})/t_1 \approx 1.1$.
In fact, while order at a HOVHS should occur for all interaction strengths~$U>0$ within mean-field approach or ladder-resummation scheme, the critical temperature drops with the fourth power of the coupling strength~\cite{Classen:2020}.
Our data suggests that the critical temperature for the formation of order at the HOVHS is already smaller than $20\,\mathrm{mK}$ due to the small interaction strength.
Nevertheless it may be possible to measure signatures of the HOVHS by scanning tunneling spectroscopy~\cite{yuan2019magic}.

\section{Functional renormalization group calculations} 

\subsection{Functional renormalization group method}

The Hartree-Fock mean-field study from the previous chapter exhibited the system's strong tendency towards the formation of magnetic order for smaller twist angles.
Notably, spin fluctuations may also provide the pairing glue for the formation of (unconventional) superconductivity, which is, however, not included in the Hartree-Fock approach.
To study competing correlated phases, including superconductivity, in the effective honeycomb-lattice Hubbard model for \tdbWSe, we therefore employ a fermionic fRG approach, which identifies the leading many-body instabilities including various types of density-wave and
superconducting instabilities on equal footing~\cite{10.1143/PTP.105.1,RevModPhys.84.299,Platt_2013,Dupuis:2020fhh}.

In our approximation, we exclusively focus on the renormalization group behavior of the normal two-particle interaction vertex~$\Gamma^{(4)}$, which is spin-rotation invariant in our model and depends on three independent wavevectors.
Schematically, the fRG introduces a scale parameter~$\Lambda$ to interpolate smoothly between the bare interactions at the ultraviolet~(UV) scale
$\Lambda=\Lambda_{\mathrm{UV}}$ and the fully dressed ones at~$\Lambda=0$. 
Our regularization scheme is based on a sharp cutoff in frequencies~\cite{PhysRevB.79.195125,RevModPhys.84.299,PhysRevResearch.6.043078} so that the vertex at scale~$\Lambda$ dresses the bare interaction with all fluctuations of energy above that scale.
In our approximation, we neglect self-energy corrections, frequency dependencies, and higher-order vertices with more than four fermionic fields. 
This truncation scheme has been frequently employed in the context of strongly-correlated electron systems  to
determine Fermi liquid instabilities in an unbiased way~\cite{10.1143/PTP.105.1,RevModPhys.84.299,PhysRevB.85.035414,Platt_2013,Dupuis:2020fhh,Profe_2024_SRO}.

A flow to strong coupling of~$\Gamma^{(4)}$ at finite $\Lambda=\Lambda_c$ signals an instability of the normal state towards a symmetry broken phase. 
Using the effective vertex at the critical scale $\Lambda_c$, we classify the instability either as a transition towards a spin/charge-density-wave state (DW) or a superconducting state (SC).
For the present system, we have employed a recent implementation of the truncated-unity FRG~\cite{Lichtenstein_2017,Profe_2022_tu2frg,Beyer_2022,Profe_2024_diverge,10.21468/SciPostPhysCodeb.26-r0.5}, which is based on a decomposition of the interaction vertex in terms of three scattering channels, i.e,  $P$, $D$ and $C$, each of which depends most strongly on the Mandelstam variables for pairing $q_P$, direct $q_D$ and crossed $q_C$ channels, respectively~\cite{PhysRevB.79.195125}.
The remaining momentum dependencies are weak and may be expanded into a truncated set of form factors that respect the symmetry of the problem at a minimal loss of accuracy~\cite{PhysRevB.79.195125,Lichtenstein_2017,Profe_2022_tu2frg}. 

For our practical fRG calculations, we made use of the divERGe library~\cite{Profe_2024_diverge,10.21468/SciPostPhysCodeb.26-r0.5}, employing the TUFRG backend with a wavevector resolution of the Mandelstam variables  with $N_q=50^2$ points in the Brillouin zone.
The wave-vector resolution in the loop integrations requires higher accuracy and we therefore choose a resolution of $N_l=2500^2$.
The truncated form-factor expansion includes all form-factor bonds connecting sites at distance equal to or smaller than $1a_M$ from either of the two sites of a reference unit cell, making up a set of eight bond form factors.

\begin{figure}
    \centering    \includegraphics[width=1.0\linewidth]{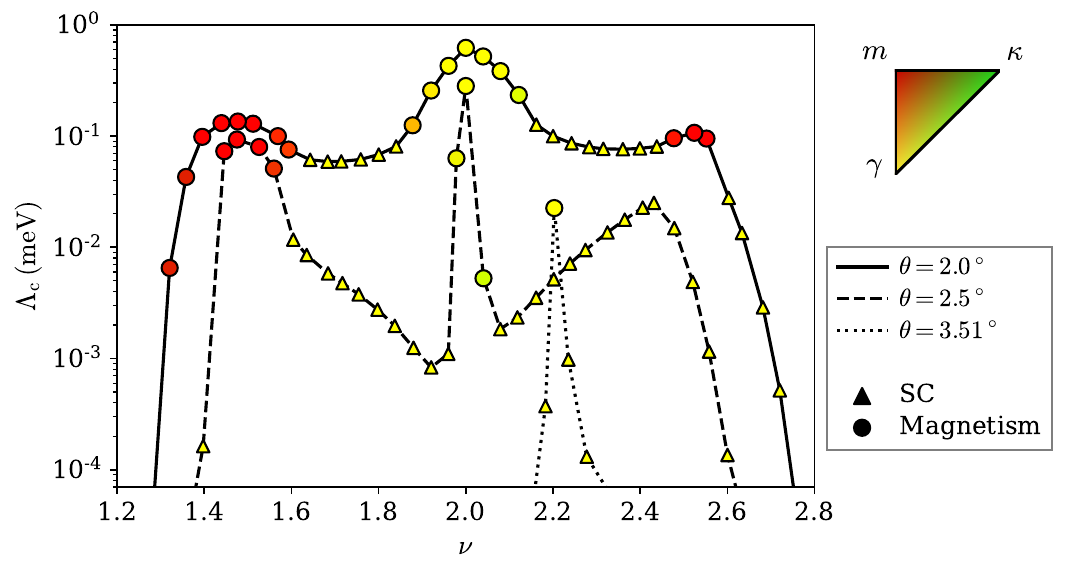}
    \caption{\textbf{fRG results for competing correlated phases.} We show results of our fRG calculations for the critical scale $\Lambda_c$ as a function of the filling~$\nu$ and for twist angles $\theta = 2.0^\circ, 2.5^\circ, 3.51^\circ$. The latter angle includes the HOVHS near filling $\nu\approx 2.2$. Closed circles indicate spin DW instabilities, whereas triangles represent SC instabilities. The color of the symbols indicates the ordering vector associated to the instability. All superconducting instabilities are found to be in the $E_2$~irreducible representation of the point group of the honeycomb lattice.}
    \label{fig:frgphasediagram}
\end{figure}

\subsection{Competing correlated phases from fRG}

To set up the numerical evaluation of the fRG for \tdbWSe, we start by gauging the interaction values~$U(\theta)$ with the experimental observation of the relativistic Mott transition occurring at the angle $\theta_0 = 2.7^\circ$ by using Eq.~\eqref{eq:hubbardUtheta} to set $U(\theta_0)$.
This is necessary because the fRG includes additional fluctuations beyond mean-field, which also tentatively mitigate the overestimation of ordering within mean-field approaches.
Indeed, we find the instability signaling this transition occurring at  $U^{\mathrm{fRG}}(\theta_0) \approx 3.4 t_1(\theta_0)$, which is closer to the numerical results from QMC~\cite{Sorella_2012,PhysRevX.3.031010,PhysRevB.101.125103} as compared to the mean-field value.

Having fixed the Hubbard interaction in this way, we now choose three representative twist angles: (i)~$\theta = 2.0^\circ$ features the approximately-nested Fermi surface at the hole-doped van Hove filling and the Hubbard interaction is already relatively strong, i.e., $U(\theta=2.0^\circ)\approx 5.5t_1$, (ii)~$\theta = 3.51^\circ$ exhibits a HOVHS at hole-doping albeit at smaller Hubbard interaction, i.e., $U(\theta=3.51^\circ)\approx 2.2t_1$, and (iii)~$\theta = 2.5^\circ$ represents an intermediate scenario where $U(\theta=2.5^\circ)\approx 3.9t_1$.
Adjusting the chemical potential, we again performed parameter scans in the filling between the VHS in the two honeycomb bands and our results are shown in Fig.~\ref{fig:frgphasediagram}.

For $\theta = 2.0^\circ$ and $\theta = 2.5^\circ$ and at Dirac filling, we find a divergence of the particle-hole exchange vertex $C$ at ordering vector~$\gamma$.
Diagonalizing the vertex $C(q = \gamma)$ in form-factor space, we find the leading eigenvector to have opposite spin projection on the two sublattices, indicating an insulating N\'eel ground state, in agreement with our Hartree-Fock study in Sec.~\ref{sec:hfphasediagram}. 
As we dope away slightly, we find an instability towards incommensurate spin DWs with momentum transfers that are also in reasonable agreement with the ones in the magnetic orders described in the mean-field analysis, cf. the middle panels of Fig.~\ref{fig:phasediagram}.
The fRG also independently provides further evidence for the tendencey towards magnetic order with momentum transfer~$m$ near the van Hove fillings, wherever this is the leading instability.

For fillings between the van Hove and the Dirac fillings we had already found in the Hartree-Fock study that the tendency towards magnetic order is somewhat weaker.
This can be seen, e.g., in the magnetization in Fig.~\ref{fig:phasediagram}, which becomes smaller or even vanishes in this region for $\theta=2.5^\circ$.
In the fRG analysis, we find that for these fillings superconducting instabilities develop, instead, see Fig.~\ref{fig:frgphasediagram}.
Diagonalization of the pairing vertex $P(q)$ in form-factor space at $q = \gamma$, identifies two degenerate leading eigenvectors that span the $E_2$ irreducible representation of the point group $C_{6v}$.
This suggests that the superconducting state is composed of a superposition of the $d_{x^2-y^2}$- and $d_{xy}$-wave form factors.
It can be further expected that a superposition of the type $d_{x^2-y^2}\pm id_{xy}$, i.e., a chiral $d$-wave superconducting state, which opens a full gap, is energetically favored~\cite{Black_Schaffer_2014}.

Remarkably, and in contrast to our observations in the Hartree-Fock analysis, we also find many-body instabilities near the HOVHS for $\theta = 3.51^\circ$.
This may be partly attributed to the slightly larger Hubbard interactions that are required to fit the experimental findings on the relativistic Mott transition. More specifically, right at the high-order van Hove filling, we find an instability in the $C$~channel at $q=\gamma$.
Further analysis of $C(q = \gamma)$ reveals the leading eigenvector to have equal spin projection on the two sublattices, indicating a ferromagnetic ground state in agreement with Ref.~\cite{Classen:2020}.
Moving the filling slightly away from the HOVHS, we find again $E_2$ superconductivity, however, with steeply decaying $\Lambda_c$. 
In the remaining region of fillings no further instabilities are found for $\theta = 3.51$.

\section{Conclusion} 

In this work, we have analyzed the $\Gamma$~valley moir\'e band structure of \tdbWSe\ with a focus on the evolution of the Fermi-surface structure of the two topmost bands upon tuning the twist angle in a range between $2^\circ$ and $4^\circ$.
An interesting feature of the twist-induced band-structure evolution is the emergence of a high-order van Hove singularity at a twist angle of $\theta_\text{HOVHS} \approx 3.58^\circ$.
We further studied the effects of electron-electron interactions in \tdbWSe\ for a material-informed effective Hubbard model using a Hartree-Fock mean-field approach in real space.

In agreement with previous studies, we find that the insulator at half filling is a N\'eel antiferromagnet, suggesting that the critical point belongs to the Gross-Neveu-Heisenberg universality class~\cite{Pan:2023,Biedermann:2025dma}.  
Away from half filling, we then identified various magnetic orders that fall into three main categories: single-mode stripe order, two orthogonal tripes, and three modes with the same amplitude. 
We find two regions close to the van Hove singularities of the upper and lower bands, where the magnetization is enhanced for twist angles~ $\lesssim 2.7^\circ$. 
In this region, the momentum transfer of the dominant spin mode lies at or close to the $m, m',m''$~points, and the order falls into the category of three modes with the same amplitudes. 
This allows for the formation of a non-coplanar SDW with non-zero spin chirality~\cite{Li_2012,PhysRevLett.101.156402} or a uniaxial SDW corresponding to a half-metal state~\cite{PhysRevLett.108.227204}.
At the high-order van Hove singularity, however, we did not find any enhancement of magnetic order within our Hartree-Fock mean-field approach, which is most likely due to an insufficient interaction strength.
On the other hand, the high-order van Hove singularity may be identified experimentally by its characteristic signal in scanning tunneling spectroscopy~\cite{yuan2019magic}.

We note that a related Hartree-Fock mean-field study of \tdbWSe\ has been very recently carried out in Ref.~\cite{Biedermann:2025dma}, directly for the continuum model with long-range Coulomb interactions. Therein, the focus lies on the case of half (or Dirac) filling and the experimentally relevant effects of uniaxial pressure or hetero-strain as well as the concomitant quantum critical behavior.
Wherever comparable, we find our results to be in very good agreement with Ref.~\cite{Biedermann:2025dma}.

As a natural extension of the Hartree-Fock analysis we carried out a fermionic functional renormalization group study of the effective honeycomb-lattice Hubbard model for \tdbWSe. 
Within the functional renormalization group approach, all electronic many-body instabilities
are treated on equal footing, which facilitates a
resolution of the competing electronic correlations. 
Importantly, this approach revealed unconventional superconductivity with a pairing gap in the $E_2$ irrep of the point group, suggesting chiral $d$-wave superconductivity. 
We furthermore find an enhancement of many-body instabilities close to the HOVHS at $\theta \approx 3.51^\circ$, which is in contrast to our findings from Hartree-Fock mean-field theory. In particular, we identify a ferromagnetic ground state right at the high-order van Hove filling, and $E_2$ superconductivity slightly away from it.

We finally note that the functional renormalization group can also be formulated in a way that explicitly includes order-parameter fluctuations~\cite{Dupuis:2020fhh}. 
This can be used to study the quantum critical fan of the Gross-Neveu-Heisenberg universality class in more detail~\cite{Tolosa-Simeon:2025fot}. 
In conclusion,
\tdbWSe{} provides an experimentally accessible platform where these theoretical predictions can be tested in a controlled way, opening a pathway towards further aspects of quantum simulation of Dirac fermions~\cite{Tolosa-Simeon:2023bqc} and beyond.

\subsection*{Acknowledgements} 

We thank Jan Biedermann, Laura Classen, Lukas Janssen, Frank Lechermann, Kin Fai Mak, and Robin Scholle for insightful discussions and useful correspondence. 
We especially thank Jonas Profe for helpful discussions and practical advice concerning the divERGe library.
BH and MMS are supported by the Mercator Research Center Ruhr under Project No.~Ko-2022-0012. 
MMS acknowledges funding from the Deutsche Forschungsgemeinschaft (DFG, German Research Foundation) under Project No.~277146847 (SFB 1238, project C02) and Project No.~452976698 (Heisenberg program).

\appendix

\section{Details on fitting procedure} \label{app:fitting}

Here, we provide more details on our fitting procedure described in Sec.~\ref{sec:effhubbard}. We aim to find the angle-dependent hopping parameters $t_n(\theta)$, such that the band structure of the  bilinear tight-binding Hamiltonian in Eq.~\eqref{eq:tbhamil}
defined on the moir\'e honeycomb superlattice imitates the two topmost bands given by the continuum model described in Sec.~\ref{sec:cont}. To that end, we fit the band structure obtained by~\eqref{eq:tbhamil} to the two topmost bands obtained from the continuum model along the one-dimensional space defined by the $\gamma$--$m$--$\kappa$--$\gamma$-path in the BZ. As noted in the main text, truthful reproduction of analytic properties of the bands from the continuum model requires an optimized choice for the number of hoppings $N_h$. 

There are various quantities that allow for a quantitative measure of a fit. 
The quantity we minimize with our fitting procedure is the residue, defined for each band as
\begin{align}
    R_n(t,E_0) = \frac{1}{N_\mathrm{path}}\! \sum_{i=1}^{N_\mathrm{path}}\! \omega_i |E_n^\mathrm{cont}(\vec{k}_i)\! -\! E_n^\mathrm{tb}(t;\vec{k}_i) \!-\! E_0|^2.
\end{align}
In the above, $E_n^\mathrm{cont}$ and $E_n^\mathrm{tb}$ denote the two topmost bands from the continuum model and the bands from the tight-binding model, respectively. Furthermore, the positive integer $N_\mathrm{path}$ denotes the number of equidistant points used to discretize the path, $\omega_i$ is a positive weight, and $E_0$ is an additional parameter used as a constant energy offset to move the tight-binding bands to the correct energy scale of the continuum bands. The weight factor $\omega_i \in [0,1]$ can be used in order to control the relevance of points $\vec{k}_i$ in the residue, i.e., if $\omega_i$ is small, the fitted band structure is allowed to deviate more from the continuum bands than at points where $\omega_i$ is larger. Here, we choose a profile where $\omega_i$ is 1 around the Dirac point $\kappa$ and the $m$-point, where we want to truthfully resolve the analytic structure of the two van Hove singularities, while points that are further away become a weight $10^{-4}$. Our fitting procedure then minimizes the sum of the residuals of the upper (subscript $u$) and lower band (subscript $l$), i.e., the function $R(t,E_0) = R_u(t,E_0) + R_l(t,E_0)$.

\begin{figure}
    \centering
    \includegraphics[width=\columnwidth]{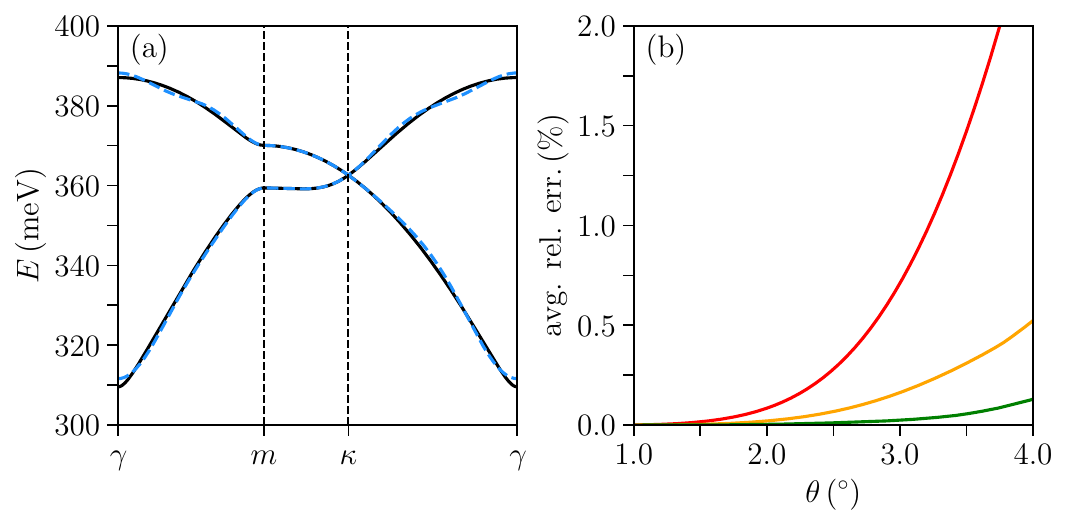}
    \caption{\textbf{Fitting procedure.} (a) Tight binding fit at $\theta=4^\circ$; (b) average relative error of the tight binding fit in respect to the continuum model for 4 (red), 6 (yellow), 10 (green) hoppings}
    \label{fig:tbfit}
\end{figure}

We observe that longer-ranged hoppings become more relevant with increasing twist angle. In Fig.~\ref{fig:tbfit}, we show the band structure for $N_h = 10$ (panel (a)) and the average relative error for $N_h = 4,6,10$ (panel (b)) for a twist angle of $4^\circ$. For the average relative error, we average over the relative error along the chosen path. We found that the best balance between analytic structure and average relative error is achieved for the choice $N_h = 10$, which is also our choice for all the results shown in the main text. We provide our optimal tight-binding parameters in Ref.~\cite{ourdata2025}.

%

\section{Details on Hartree-Fock scheme} \label{app:hfmft}

Here, we provide more details on our numerical implementation of the Hartree-Fock mean-field scheme used to produce the phase diagrams in the main text. 
Within the Hartree-Fock mean-field approach, the Hubbard interaction is replaced by the bilinear Hamiltonian given in Eq.~\eqref{eq:HMFT}.
The full mean-field Hamiltonian is then given by
\begin{equation}
    H^{\mathrm{MF}} = H_0 + H_{\mathrm{int}}^{\mathrm{MF}} = \sum_{j,j^\prime} \sum_{\sigma\sigma^\prime} c_{j\sigma}^\dagger \mathcal{H}_{jj^\prime}^{\sigma\sigma^\prime} c_{j^\prime\sigma^\prime} + \mathrm{const.}\, ,
\end{equation}
where $\mathcal{H}_{jj^\prime}^{\sigma\sigma^\prime}$ is a $2N \times 2N$--dimensional single-particle Hamiltonian. 
We denote the eigenvalues and eigenvectors of $(\mathcal{H}_{jj^\prime}^{\sigma\sigma^\prime})$ as $\epsilon_l$ and $v_{i\sigma}^l$ with $l=1,...,2N$, respectively.

For a given set of mean-field parameters, the free energy density is given by 
\begin{align}
    \frac{F}{N} = &- \frac{T}{N} \sum_l \ln\left(1+e^{-\beta (\epsilon_l - \mu)}\right) + \mu n \nonumber\\ 
    &-\frac{1}{NU_0}\sum_j \left(\Delta_{j\uparrow}\Delta_{j\downarrow} - \Delta_{j-}\Delta_{j+} \right),\label{eq:fenergy}
\end{align}
where $\mu$ is the chemical potential.\\[2pt]

Our Hartree-Fock scheme consists of the following steps:

\begin{figure*}[t!]
    \centering
    \includegraphics[width=\textwidth]{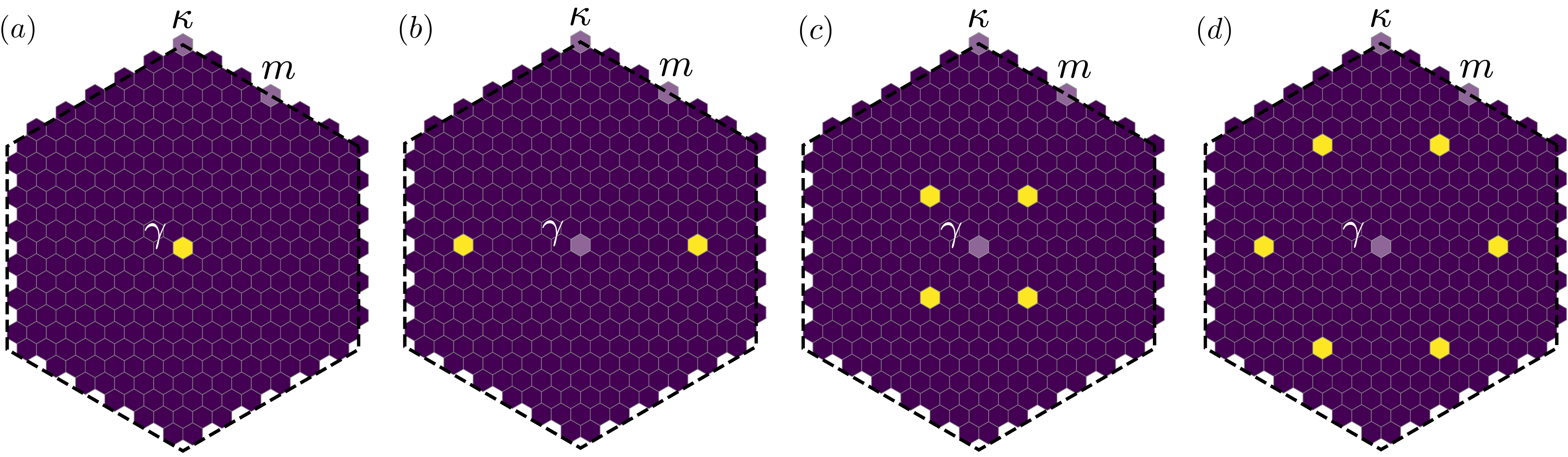}
    \caption{\textbf{Momentum-space pattern of dominant modes $\vec{S}_{\vec{q}}$.} We show exemplary wavevectors $\vec{q}$ of dominant modes $\vec{S}_{\vec{q}}$ (yellow hexagons) corresponding to (a) collinear magnetism, (b) single-mode stripe order, (c) two orthogonal stripes, and (d) three modes with same amplitude, in the hexagonal BZ.}
    \label{fig:momspacepat}
\end{figure*}

\begin{enumerate}
    \item We start the Hartree-Fock iteration with randomly generated mean-field parameters that are subsequently rescaled such that $\sum_{j\sigma} \langle n_{j\sigma} \rangle/N = n$.
    \item We calculate the eigenvalues and -vectors of $(\mathcal{H}_{jj^\prime}^{\sigma\sigma^\prime})$.
    \item Since the band structure changes with each Hartree-Fock iteration step, the chemical potential has to be adjusted to maintain constant filling. This is achieved by determining $\mu$ such that 
    \begin{align}
    n = \frac{1}{N} \sum_{l=1}^{N} f(\epsilon_l-\mu), \ f(x)=\frac{1}{e^{x/k_B T}+1}.
    \end{align}
    A solution of the above is a zero of the function
    \begin{align}
        g(\mu) = n - \frac{1}{N} \sum_{l=1}^{2N} f(\epsilon_l - \mu).
    \end{align}
    Note that the function $g(\mu)$ is monotonically decreasing, that is, $dg/d\mu(\mu) \leq 0$ for all $\mu$, where equality is only possible at zero temperature. For finite filling, it holds that $\mu \in [\mu_1,\mu_2] = [\min_l{\epsilon_l}, \max_l{\epsilon_l}]$. On the boundary, we find
    \begin{align}
        g(\mu_1) > n - 2, \quad g(\mu_2) < n - 2.
    \end{align}
    Thus, the unique zero of $g(\mu)$ lies in the interval $[\mu_1,\mu_2]$, which can be conveniently found with a bisection method, such as Brent's method.
    
    \item Once the correct chemical potential is determined, the mean-field parameters can be updated by calculating the expectation values
    \begin{equation}
        \langle c_{i\sigma}^\dagger c_{j\sigma^\prime} \rangle = \sum_{l=1}^{N} (v_{i\sigma}^l)^* v_{j\sigma^\prime}^l f(\epsilon_l - \mu).
    \end{equation}
    \item We now compare the newly calculated mean-field parameters with those of the previous iteration. If the error is below a certain threshold, which we chose to be $1\cdot 10^{-9}$, or 10000 iterations are reached, the Hartree-Fock scheme is considered as converged. Otherwise, the above steps are repeated with  $(\mathcal{H}_{jj^\prime}^{\sigma\sigma^\prime})$ containing the updated parameters. To improve the convergence, we employ a linear mixing between each iteration, i.e., we use $70\%$ of the current and $30\%$ of the previous expectation values.
\end{enumerate}

After each Hartree-Fock calculation, we determine the free energy density according to Eq.~\eqref{eq:fenergy}. To ensure that we did not converge into a local minimum, we repeat the Hartree-Fock calculation three times starting from different random initial conditions, compare their free energies, and finally identify the ground state as the result with the lowest free energy.

\section{Classification of magnetic states} \label{app:class}

Here, we follow Refs.~\cite{Scholle2023hubbard,scholle2024mean} and provide a classification of different stripe orders based on the Fourier expansion in Eq.~\eqref{eq:spinfourier}. A magnetic state with $m > 0$ can be classified by analyzing the modes $\vec{S}_{\vec{q}}$ appearing in the Fourier transform
\begin{equation} 
\label{eqapp:fourier}
    \langle \vec{S}_i \rangle = \sum_{\vec{q}} \vec{S}_{\vec{q}} e^{i \vec{q} \cdot \vec{R}_i},
\end{equation}
where $\vec{q}$ runs over the BZ of the moire reciprocal lattice.

A paramagnetic state in the thermodynamical limit is uniquely classified by a vanishing magnetization, i.e., $m = 0$, which is equivalent to $\langle \vec{S}_i \rangle = 0$. In terms of the Fourier coefficients in Eq.~\eqref{eqapp:fourier}, a state is paramagnetic if $\vec{S}_{\vec{q}} = 0$ for all $\vec{q}$ in the Brillouin zone due to the fact that the Fourier expansion is an involution. Note, however, that on a finite lattice,  magnetization $m$ and expectation value $\langle \vec{S}_i \rangle$ can be non-vanishing even in the paramagnetic phase. Hence, a sharp transition into a magnetically ordered phase never occurs due to the analyticity of the free energy. Therefore, we choose a criterion on the basis of which we classify a state as paramagnetic, even if the magnetization is small but finite. Here, we follow Ref.~\cite{Scholle2023hubbard}, and choose a threshold based on the Fourier modes, i.e., we consider a state as paramagnetic if
\begin{align}
    \frac{1}{N} \sqrt{\sum_{\vec{q}} |\vec{S}_{\vec{q}}|^2} < a_\mathrm{tol}.
\end{align}
For the phase diagrams shown in the main text, we choose $a_\mathrm{tol}$ such that our results for $T = 20\, \mathrm{mK}$ at half filling are compatible with the experimental findings, i.e., that we are in the paramagnetic phase for $\theta > \theta_c = 2.7^\circ$, cf. Sec.~\ref{sec:relmott}. This yields $a_\mathrm{tol} = 3 \cdot 10^{-2}.$

Similar to Refs.~\cite{Sachdev2019stripe,Millis2022,Scholle2023hubbard,scholle2024mean}, we further distinguish different stripe orders as follows.

\textit{Single-mode stripe order:}
In the case of single-mode stripe order, only one mode with $\vec{q} = \pm \vec{Q}$ appears in the Fourier expansion, i.e.,
\begin{equation}
     \vec{S}_{\vec{Q}} = \vec{S}_{-\vec{Q}}^* = \mathcal{S} \begin{pmatrix} 1 \\ 0 \\ 0 \end{pmatrix}, \qquad \vec{S}_{\vec{q}} = \vec{0} \quad \mathrm{for} \quad \vec{q} \neq \pm\vec{Q}.
\end{equation}

\textit{Two orthogonal stripes:} Two orthogonal stripe order is characterized by two orthogonal single-stripe modes, i.e.,
\begin{align}
    \vec{S}_{\vec{Q}_1} = \vec{S}_{-\vec{Q}_1}^* = \mathcal{S}_1 \begin{pmatrix} 1 \\ 0 \\ 0 \end{pmatrix},\quad 
    \vec{S}_{\vec{Q}_2} = \vec{S}_{-\vec{Q}_2}^* = \mathcal{S}_2 \begin{pmatrix} 0 \\ 1 \\ 0 \end{pmatrix}.
\end{align}

\textit{Three modes with same amplitude:} This state is classified by $S_{\vec{q}}$ being nonzero only for three modes $\pm\vec{Q}_1,\pm\vec{Q}_2,\pm\vec{Q}_3$ that have the same amplitude,
\begin{equation}
    |\vec{S}_{\vec{Q}_1} | = | \vec{S}_{\vec{Q}_2} | = | \vec{S}_{\vec{Q}_3} |.
\end{equation}
This class contains the following two special cases.

\textit{Orthogonal stripes:}
The three modes with the same amplitude are orthogonal to each other, i.e.
\begin{align}
    \vec{S}_{\vec{Q}_1} = \mathcal{S} \begin{pmatrix} 1 \\ 0 \\ 0 \end{pmatrix},\quad
    \vec{S}_{\vec{Q}_2} = \mathcal{S} \begin{pmatrix} 0 \\ 1 \\ 0 \end{pmatrix},\quad
    \vec{S}_{\vec{Q}_3} = \mathcal{S} \begin{pmatrix} 0 \\ 0 \\ 1 \end{pmatrix}.
\end{align}

\textit{Collinear stripes:}
Three spin amplitudes are collinear, i.e.,
\begin{equation}
    \vec{S}_{\vec{Q}_1} = \vec{S}_{\vec{Q}_2} =  \vec{S}_{\vec{Q}_3} = \mathcal{S} \begin{pmatrix} 1 \\ 0 \\ 0 \end{pmatrix}.\\
\end{equation}

\textit{Other order:}
If a spin order does not fall into one of the orders mentioned above, we categorize it as ``other''.

We show the momentum-space pattern of the dominant wavevectors of the different orders in Fig.~\ref{fig:momspacepat}. Note that the different ordered states will fulfill the above conditions exactly only in the true thermodynamic limit. On a lattice with a finite amount of sites, such classification is, however, still possible among the `dominant' Fourier modes $\vec{S}_{\vec{q}}$. Here, we define dominant modes as a set of modes with descending norm, which have a weight of $95\%$ of the total weight.

\bibliographystyle{longapsrev4-2}
\bibliography{references}

\end{document}